\documentclass[11pt]{article}
\usepackage{amsmath,amsthm,latexsym,amssymb,amsfonts,epsfig}

\DeclareMathOperator{\MC}{\boldsymbol{\mathsf{M}}}
\DeclareMathOperator{\MCO}{\boldsymbol{\widehat{\mathsf{M}}}}
\DeclareMathOperator{\MA}{\boldsymbol{\mathsf{W}}}

\DeclareMathOperator{\MCPW}{\boldsymbol{\mathsf{Master\;\;Constraint\;\;
Programme}}}

\DeclareMathOperator{\TPPW}{\boldsymbol{\mathsf{The 
\; Phoenix\;\;Project:}}}
\DeclareMathOperator{\PPW}{\boldsymbol{\mathsf{Phoenix\;\;Project}}}
\DeclareMathOperator{\KD}{\boldsymbol{\mathsf{K}}}

\makeatletter
\@addtoreset{equation}{section}
\makeatother

\newtheorem{Theorem}{Theorem}[section]
\newtheorem{Definition}{Definition}[section]
\newtheorem{Lemma}{Lemma}[section]

\def\be{\begin{equation}}
\def\ee{\end{equation}}
\def\ba{\begin{eqnarray}}
\def\ea{\end{eqnarray}}
\def\a{{\cal A}}
\def\ab{\overline{\a}}

\def\Nl{{\mathchoice
{\setbox0=\hbox{$\displaystyle\rm N$}\hbox{\hbox to0pt
{\kern0.4\wd0\vrule height0.9\ht0\hss}\box0}}
{\setbox0=\hbox{$\textstyle\rm N$}\hbox{\hbox to0pt
{\kern0.4\wd0\vrule height0.9\ht0\hss}\box0}}
{\setbox0=\hbox{$\scriptstyle\rm N$}\hbox{\hbox to0pt
{\kern0.4\wd0\vrule height0.9\ht0\hss}\box0}}
{\setbox0=\hbox{$\scriptscriptstyle\rm N$}\hbox{\hbox to0pt
{\kern0.4\wd0\vrule height0.9\ht0\hss}\box0}}}}
\def\Zl{{\mathchoice
{\setbox0=\hbox{$\displaystyle\rm Z$}\hbox{\hbox to0pt
{\kern0.4\wd0\vrule height0.9\ht0\hss}\box0}}
{\setbox0=\hbox{$\textstyle\rm Z$}\hbox{\hbox to0pt
{\kern0.4\wd0\vrule height0.9\ht0\hss}\box0}}
{\setbox0=\hbox{$\scriptstyle\rm Z$}\hbox{\hbox to0pt
{\kern0.4\wd0\vrule height0.9\ht0\hss}\box0}}
{\setbox0=\hbox{$\scriptscriptstyle\rm Z$}\hbox{\hbox to0pt
{\kern0.4\wd0\vrule height0.9\ht0\hss}\box0}}}}
\def\Ql{{\mathchoice
{\setbox0=\hbox{$\displaystyle\rm Q$}\hbox{\hbox to0pt
{\kern0.4\wd0\vrule height0.9\ht0\hss}\box0}}
{\setbox0=\hbox{$\textstyle\rm Q$}\hbox{\hbox to0pt
{\kern0.4\wd0\vrule height0.9\ht0\hss}\box0}}
{\setbox0=\hbox{$\scriptstyle\rm Q$}\hbox{\hbox to0pt
{\kern0.4\wd0\vrule height0.9\ht0\hss}\box0}}
{\setbox0=\hbox{$\scriptscriptstyle\rm Q$}\hbox{\hbox to0pt
{\kern0.4\wd0\vrule height0.9\ht0\hss}\box0}}}}
\def\Rl{{\mathchoice
{\setbox0=\hbox{$\displaystyle\rm R$}\hbox{\hbox to0pt
{\kern0.4\wd0\vrule height0.9\ht0\hss}\box0}}
{\setbox0=\hbox{$\textstyle\rm R$}\hbox{\hbox to0pt
{\kern0.4\wd0\vrule height0.9\ht0\hss}\box0}}
{\setbox0=\hbox{$\scriptstyle\rm R$}\hbox{\hbox to0pt
{\kern0.4\wd0\vrule height0.9\ht0\hss}\box0}}
{\setbox0=\hbox{$\scriptscriptstyle\rm R$}\hbox{\hbox to0pt
{\kern0.4\wd0\vrule height0.9\ht0\hss}\box0}}}}
\def\Cl{{\mathchoice
{\setbox0=\hbox{$\displaystyle\rm C$}\hbox{\hbox to0pt
{\kern0.4\wd0\vrule height0.9\ht0\hss}\box0}}
{\setbox0=\hbox{$\textstyle\rm C$}\hbox{\hbox to0pt
{\kern0.4\wd0\vrule height0.9\ht0\hss}\box0}}
{\setbox0=\hbox{$\scriptstyle\rm C$}\hbox{\hbox to0pt
{\kern0.4\wd0\vrule height0.9\ht0\hss}\box0}}
{\setbox0=\hbox{$\scriptscriptstyle\rm C$}\hbox{\hbox to0pt
{\kern0.4\wd0\vrule height0.9\ht0\hss}\box0}}}}
\def\Hl{{\mathchoice
{\setbox0=\hbox{$\displaystyle\rm H$}\hbox{\hbox to0pt
{\kern0.4\wd0\vrule height0.9\ht0\hss}\box0}}
{\setbox0=\hbox{$\textstyle\rm H$}\hbox{\hbox to0pt
{\kern0.4\wd0\vrule height0.9\ht0\hss}\box0}}
{\setbox0=\hbox{$\scriptstyle\rm H$}\hbox{\hbox to0pt
{\kern0.4\wd0\vrule height0.9\ht0\hss}\box0}}
{\setbox0=\hbox{$\scriptscriptstyle\rm H$}\hbox{\hbox to0pt
{\kern0.4\wd0\vrule height0.9\ht0\hss}\box0}}}}
\def\Ol{{\mathchoice
{\setbox0=\hbox{$\displaystyle\rm O$}\hbox{\hbox to0pt
{\kern0.4\wd0\vrule height0.9\ht0\hss}\box0}}
{\setbox0=\hbox{$\textstyle\rm O$}\hbox{\hbox to0pt
{\kern0.4\wd0\vrule height0.9\ht0\hss}\box0}}
{\setbox0=\hbox{$\scriptstyle\rm O$}\hbox{\hbox to0pt
{\kern0.4\wd0\vrule height0.9\ht0\hss}\box0}}
{\setbox0=\hbox{$\scriptscriptstyle\rm O$}\hbox{\hbox to0pt
{\kern0.4\wd0\vrule height0.9\ht0\hss}\box0}}}}

\title{$\TPPW$ \\
$\MCPW$\\ for Loop Quantum Gravity}
\author{T. Thiemann\thanks{thiemann@aei-potsdam.mpg.de}~\thanks{New 
address: Perimeter Institute for Theoretical Physics and University of 
Waterloo, 35 Kingt Street North, Waterloo, Ontario N2J 2G9, Canada, 
email: tthiemann@perimeterinstitute.ca},
       MPI f. Gravitationsphysik, Albert-Einstein-Institut, \\
           Am M\"uhlenberg 1, 14476 Golm near Potsdam, Germany}
\date{{\small Preprint AEI-2003-047, PI-2003-003}}

\begin{document}

\maketitle

\begin{abstract}
The Hamiltonian constraint remains the major unsolved problem in Loop
Quantum Gravity (LQG). Seven years ago a mathematically consistent 
candidate Hamiltonian constraint has been proposed but there are 
still several unsettled questions which concern the algebra of commutators 
among smeared Hamiltonian constraints which must be faced in 
order to make progress. 

In this paper we propose a solution to this set of problems based 
on the so-called {\bf Master Constraint} which combines the smeared 
Hamiltonian constraints for all smearing functions into a single 
constraint. If certain mathematical conditions, which still have 
to be proved, hold, then not only the problems with the commutator algebra 
could disappear, also chances are good that one can control the solution 
space and the (quantum) Dirac observables of LQG. Even a decision
on whether the theory has the correct classical limit
and a connection with the path integral (or spin foam) formulation
could be in reach.

While these are exciting possibilities, we should warn the reader from 
the outset that, since the proposal is, to 
the best of our knowledge, completely new and has been barely tested 
in solvable models, there might be caveats which we are presently unaware 
of and render the whole {\bf Master Constraint Programme} obsolete. Thus,
this paper should really be viewed as a proposal only, rather than a 
presentation of hard results, which however we intend to supply in future
submissions. 
\end{abstract}

\section{Introduction}
\label{s1}

The quantum dynamics has been the most difficult technical and conceptual 
problem for quantum gravity ever since. This is also true for Loop
Quantum Gravity (LQG) (see \cite{1} for recent reviews). Seven years 
ago, for the first time a mathematically well-defined Hamiltonian
constraint operator has been proposed 
\cite{2,3,4,5,6,7,8,9} for LQG which is a candidate for the definition 
of the quantum dynamics of the gravitational field and all known 
(standard model) matter. That this is 
possible at all is quite surprising because the Hamiltonian constraint
of classical General Relativity (GR) is a highly non-polynomial function 
on phase space and the corresponding operator should therefore be plagued 
with UV singularities even more serious than for interacting, 
ordinary Quantum Field Theory. In fact, it should be at this point 
where the non-renormalizability of perturbative quantum (super)gravity
is faced in LQG \cite{10}. That this does not happen is a direct 
consequence of the 
background independence of LQG which is built in at a fundamental level
and therefore requires a non-perturbative definition of the theory in 
a representation \cite{11,12,13,14} which is fundamentally different 
from the usual background dependent Fock representations.

Despite this success, immediately after the the appearance of 
\cite{2,3,4,5,6,7,8,9} three papers \cite{15,16,17} were 
published which criticized the proposal by doubting the correctness of the 
classical limit of the Hamiltonian constraint operator. In broad terms,
what these papers point out is that, 
while the algebra of commutators among smeared Hamiltonian constraint
operators is anomaly free in the mathematical sense (i.e. does not not
lead to inconsistencies), it does not manifestly reproduce the classical
Poisson algebra among the smeared Hamiltonian constraint functions. While 
the arguments put forward are inconclusive (e.g. the direct translation 
of the techniques used in the full theory work extremely well in Loop
Quantum Cosmology \cite{18}) these three papers raised a 
serious issue and presumably discouraged almost all researchers in the 
field to
work on an improvement of these questions. In fact, except for two
papers \cite{19} there has been no publication on possible modifications
of the Hamiltonian constraint proposed. Rather, the combination of 
\cite{2,3,4} with path integral techniques \cite{20} and ideas from 
topological quantum field theory \cite{21} gave rise to the so-called 
spin foam reformulation of LQG \cite{22} (see e.g. \cite{23} for a recent
review). Most of the activity in LQG over the past five years has 
focussed on spin foam models, partly because the hope was that spin foam 
models, which are defined rather independently of the Hamiltonian 
framework, circumvent the potential problems pointed out in 
\cite{15,16,17}. However, the problem reappears as was shown in recent
contributions \cite{24} which seem to indicate that the whole virtue of 
the spin foam formulation, its manifestly covariant character, does not 
survive quantization. 

One way out could be to look at constraint quantization from an entirely
new point of view \cite{25} which proves useful also in discrete 
formulations of classical GR, that is, numerical GR. While being a 
fascinating possibility, such a procedure would be a 
rather drastic step in the sense that it would render most results of LQG 
obtained so far obsolete.\\ 
\\
\\
In this paper we propose a new, more modest, method to {\bf cut the Gordic 
Knot} which we will describe in detail in what follows. Namely we 
introduce the $\PPW$ which aims at {\it reviving interest in the 
quantization of the Hamiltonian Constraint}. However, before the reader 
proceeds we would like to express a word of warning:\\ 
So far this is really only a proposal. While there are many promising 
features as we will see, many mathematical issues, mostly functional 
analytic in nature, are not yet worked out completely. Moreover, the 
proposal is, to the best of our 
knowledge, completely new and thus has been barely tested in solvable 
models. Hence, there might be possible pitfalls which we are simply 
unaware of at present and which turn the whole programme obsolete.
On the other hand there are so many mathematical facts which work together
harmonically that it would be a pity if not at least part of our idea 
is useful. It is for this reason that we dare to publish this paper 
although the proposal is still premature. This should 
be kept in mind for waht follows.\\
\\
\\
The origin of the potential problems pointed out in \cite{15,16,17}
can all be traced back to simple facts about the constraint algebra:\\ 
\\
{\it 1. The (smeared) Hamiltonian constraint is not a spatially 
diffeomorphism invariant function.\\
2. The algebra of (smeared) Hamiltonian constraints does not close,
it is proportional to a spatial diffeomorphism constraint.\\
3. The coefficient of proportionality is not a constant, it is a 
non-trivial function on phase space whence the constraint algebra is open 
in the BRST sense.}\\ 
\\
These phrases are summarized in the well-known formulas (Dirac or 
hpersurface deformation algebra)
\ba
\{\vec{C}(\vec{N}),\vec{C}(\vec{N}')\} &=& 
\kappa \vec{C}({\cal L}_{\vec{N}}\vec{N}')
\nonumber \\
\{\vec{C}(\vec{N}),C(N')\} &=& 
\kappa C({\cal L}_{\vec{N}} N')
\nonumber \\
\{C(N),C(N')\} &=&\kappa
\int_\sigma d^3x (N_{,a} N'-N N'_{,a})(x) q^{ab}(x) C_b(x)
\nonumber
\ea
where $C(N)=\int_\sigma d^3x N(x) C(x)$ is the smeared Hamiltonian 
constraint, $C_b$ is the spatial diffeomorphism constraint, 
$\vec{C}(\vec{N})=\int_\sigma d^3x N^a(x) C_a(x)$ is the smeared 
spatial diffeomorphism constraint,
$q^{ab}$ is the 
inverse spatial metric tensor, $N,N',N^a,N^{\prime a}$ are smearing 
functions on the spatial three-manifold $\sigma$ and $\kappa$ is the 
gravitational constant.

This is actually the source of a whole bunch of difficulties which 
make the regularization of the Hamiltonian constraint a delicate issue
if one wants to simultaneously avoid an inconsistency of the constraint 
algebra. 
Notice that due to the third relation the Dirac algebra is fantastically
much more complicated than any infinite dimensional Lie (Super)algebra.
Moreover, while
the Hamiltonian constraint of \cite{2,3,4} uses spatial diffeomorphism 
invariance in an important way in order to remove the regulator in the 
quantization procedure, the resulting operator does not act on spatially
diffeomorphism invariant states (it maps diffeomorphim invariant
states to those which are not) thus preventing us from using the Hilbert 
space of spatially diffeomorphism states constructed in \cite{26}. 

The observation of this paper is that all of this would disappear if it 
would be possible to reformulate the Hamiltonian constraint in such a 
way that it is equivalent to the original formulation but such that it
becomes a spatially diffeomorphism invariant function with an honest
constraint Lie algebra. There is a natural candidate, namely
$$
\MC=\int_\sigma d^3x \frac{[C(x)]^2}{\sqrt{\det(q(x))}}
$$
We call it the {\bf Master Constraint} corresponding to the infinite 
number of constraints $C(x),\;x\in\sigma$ because, due to positivity 
of the integrand, the {\bf Master Equation} $\MC=0$ is equivalent 
with $C(x)=0\;\;\forall x\in\sigma$ since $C(x)$ is real valued. (One 
could also consider higher, positive powers of $C(x)$ but quadratic 
powers are the simplest). The factor $1/\sqrt{\det(q)}$ has been 
incorporated in order to 
make the integrand a scalar density of weight one (remember that 
$C(x)$ is a density of weight one). This guarantees 1) that $\MC$
is a spatially diffeomorphism invariant quantity and 2) that 
$\MC$ has a chance to survive quantization \cite{7}. 

Why did one not 
think of such a quantity before ? In fact, related ideas have been
expressed already: In \cite{27} the authors did construct 
an {\it infinite number} of modified Hamiltonian constraints $\KD(x)$,
rather than a {\it single} {\bf Master Constraint}, which have an 
Abelean algebra among themselves. One can show that none of these {\bf 
Kucha\v{r} Densities} $\KD(x)$ is simulataneously 1) a density of weight 
one, 2) a 
polynomial in $C(x)$ and 3) positive definite. Rather, they are 
algebraic aggregates built from $C(x)^2$ and $C(x)^2-(q^{ab} C_a C_b)(x)$.
If it is not a polynomial, then it will be not differentiable on the 
constraint surface and if it is not a density of weight one then it cannot 
be quantized background independently. Thus from this point of view, $\MC$ 
is an improvement, since clearly $\{\MC,C_a(x)\}= 
\{\MC,\MC\}=0$, moreover the number of constraints is drastically reduced. 
But still there is 
an a priori problem with $\MC$ which prevented the author from 
considering it seriously much earlier: On the constraint surface 
$\MC=0$ we obviously have $\{O,\MC\}=0$ for {\it any}
differentiable function $O$ on the phase space. This is a problem 
because (weak) Dirac observables for first class constraints such as 
$C(x)=0$ are selected precisely by the condition $\{O,C(x)\}=0$ for 
all $x\in \sigma$ on the constraint surface. Thus the {\bf Master 
Constraint} seems to fail to 
detect Dirac observables with respect to the original set of Hamiltonian 
constraints $C(x)=0,\;x\in\sigma$. 

The rather trivial, but yet important 
observation is that this is {\it not the case}: We will prove that an at 
least twice differentiable function $O$ on phase space is a weak Dirac 
observable with respect to all Hamiltonian constraints $C(x)=0,\;
x\in\sigma$ if and only if it satisfies the single {\bf Master Equation}
$$
\{O,\{O,\MC\}\}_{\MC=0}=0
$$
The price we have to pay in order to replace the infinite number 
of linear (in $O$) conditions $\{O,C(x)\}_{\MC=0}=0,
\;x\in\sigma$ by this single {\bf Master
Equation} is that it becomes a non-linear condition on $O$. This is 
a mild price to pay in view of having only a single equation to solve.
Now from the theory of differential equations one knows that non-linear
partial differential equations (such as the Hamilton-Jacobi equation)
are often easier to solve if one transforms them first into a system 
of linear (ordinary, in the case of Hamilton's equations,) partial 
differential equations and one might think that in order to find solutions 
to the {\bf Master Equation} one has to go back to the original infinite 
system of conditions. However, also this is not the case: As we will 
show, one can 
explicitly solve the {\bf Master Condition} for the subset of {\it strong}
Dirac observables by using {\bf Ergodic Theory Methods}. 

Hence, the {\bf Master Constraint} seems to be quite useful at the 
classical level. How about the quantum theory ? It is here where things 
become even more beautiful: Several facts {\it work together 
harmonically}:
\begin{itemize}
\item[i)] {\it Spatially Diffeomorphism Invariant States}\\
The complete space of solutions to the spatial diffeomorphism constraints
$C_a(x)=0,\;x\in\sigma$ has already been found long ago in \cite{26} and 
even was equipped with a natural inner product induced from that of the 
kinematical Hilbert space ${\cal H}_{Kin}$ of solutions to the Gauss 
constraint.
However, there is no chance to define the Hamiltonian constraint 
operators corresponding to $C(x)$ (densely) on ${\cal H}_{Diff}$ because 
the Hamiltonian constraint operators do not preserve ${\cal H}_{Diff}$.
The Hamiltonian constraints $C(x)$ therefore had to be defined on 
${\cal H}_{Kin}$ 
but this involves the axiom of choice and thus introduces a huge quantization
ambiguity \cite{3}. Moreover, removal of the regulator of the regulated 
constraint operator is possible only in an unusual operator topology which 
involves diffeomorphism invariant distributions.

However, $\MC$ is spatially diffeomorphism invariant and therefore {\it 
can be defined on} ${\cal H}_{Diff}$. Therefore, we are finally able to
{\bf exploit the full power of the results obtained in \cite{26}}! Hence 
the ambiguity mentioned disappears, the convergence of the regulated 
operator is the standard weak operator topology of ${\cal H}_{Diff}$, the 
whole quantization becomes much cleaner. Yet, all the steps carried out 
in \cite{2,3,4,5,6,7,8} still play an important role for the quantization
of $\MC$, so these efforts were not in vain at all.
\item[ii)] {\it Physical States and Physical Inner Product}\\
What we are a priori constructing is actually not an operator corresponding 
to $\MC$ but rather only a (spatially diffeomorphism invariant) quadratic 
form $Q_{\MC}$ on ${\cal H}_{Diff}$. That is, we are able to compute the 
matrix 
elements of the would-be operator $\MCO$. For most
practical purposes this is enough, however, things become even nicer if 
$Q_{\MC}$ really is induced by an actual operator on 
${\cal H}_{Diff}$. Now {\it by construction the quadratic form 
$Q_{\MC}$ is positive}. However, for semi-bounded quadratic
forms, of which the positive ones are a subset, general theorems in 
functional analysis guarantee that there is a unique, positive, 
self-adjoint operator $\MCO$ whose matrix elements 
reproduce $Q_{\MC}$, provided that the quadratic form is 
closable. Now, although we do not have a full proof yet, since 
$Q_{\MC}$ is not some random positive quadratic form but 
actually comes from a positive function on the phase space, chances are 
good that a closed extension exists. 

Let us assume that this is actually the case. Now in contrast to 
${\cal H}_{Kin}$ the Hilbert space ${\cal H}_{Diff}$ {\it is separable}.
(This is not a priori the case but can be achieved by a minor 
modification of the procedure in \cite{26}). It is, in general, only for 
{\it separable}
Hilbert spaces that the following theorem holds: There is a direct integral
decomposition of ${\cal H}_{Diff}$ corresponding to the self-adjoint 
operator $\MCO$ into 
Hilbert 
spaces ${\cal H}^\oplus_{Diff}(\lambda),\;\lambda\in\Rl$ such that the action
of $\MCO$ on ${\cal H}^\oplus_{Diff}(\lambda)$ 
reduces to multiplication by $\lambda$. Hence, {\it the physical Hilbert 
space} is simply given by 
${\cal H}_{Phys}={\cal H}^\oplus_{Diff}(0)$. Notice that this Hilbert
space {\it automatically comes with its own physical inner product} which 
is induced by that of ${\cal H}_{Diff}$ (which in turn is induced by that
of ${\cal H}_{Kin}$). So we would have shown automatically existence of 
${\cal H}_{Phys}$, however, it is not a priori clear if it is 
sufficiently large (contains enough semiclassical states). That is, 
while the constraint algebra with respect to $\MCO$ has been trivialized,
operator ordering choices still will play an important role in the 
sense that they will have influence on the size of ${\cal H}_{Phys}$.
Hence, the issue of {\bf anomaly freeness} has been transformed into
the issue of {\bf the size of ${\cal H}_{Phys}$}. Hence it seems that 
nothing has been gained, but this is not true: The {\bf Master Constraint 
Method} allows us to postpone operator issues until the very end of the 
analysis with the advantage that there are no additional mathematical 
obstacles along the way. 
\item[iii)] {\it Strong Quantum Dirac Observables}\\
Again, if a self-adjoint operator $\MCO$ exists, then 
we can construct the weakly continuous one-parameter unitary groups 
$t\mapsto \hat{U}(t)$ generated by it. Let $\hat{O}_{Diff}$ be a bounded, 
self-adjoint operator on ${\cal H}_{Diff}$, say one of the spectral 
projections of a normal operator on ${\cal H}_{Diff}$. Suppose that 
$$
[\hat{O}]:=\lim_{T\to\infty} \frac{1}{2T} \int_{-T}^T\; dt\;
\hat{U}(t)\;\hat{O}_{Diff}\; \hat{U}(t)^{-1}
$$
converges in the uniform operator topology to a bounded, symmetric
operator on ${\cal H}_{Diff}$ (and can thus be extended to a self-adjoint
operator there). 
Then (the spectral projections of) $[\hat{O}]$ commute(s) with (the 
spectral projections of) $\MCO$, hence it leaves 
${\cal H}_{Phys}$ invariant and  
induces a bounded, symmetric, hence self adjoint operator $\hat{O}_{Phys}$ 
there. 
Then this {\it ergodic mean technique} would be a simple tool in order to
construct {\it strong quantum Dirac Observables} with the correct
(induced) adjountness relations.
\end{itemize}
In summary, we feel that 1. {\bf Diffeomorphism Invariance of} $\MC$, 2.
{\bf Positivity of} $\MC$ and 3. {\bf Separability} of ${\cal H}_{Diff}$ 
work 
together harmonically and provide us with powerful functional analytic
tools which are not at our disposal when working with the Hamiltonian 
constraints.\\
\\
This finishes the introduction. This is the first of a series of papers
in which we will analyze the functional analytic properties of the 
{\bf Master Constraint Operator} and hopefully complete all the missing 
steps in our programme. This first paper just aims at sketching the 
broad outlines of that programme, more details will follow in subsequent 
submissions. We have made an effort to keep the number of formulas 
and theorems at a minimum while keeping the paper self-contained so that 
also theoretical and mathematical physicists with a surfacial knowledge 
of LQG can access the paper. More details and proofs for the experts will 
hopefully follow in future submissions.

The article breaks up into the following sections.\\ 
\\

In section two we compare the 
{\bf Master Constraint Programme} applied to General Relativity with
the original Hamiltonian 
Constraint Programme of \cite{2,3,4,5,6,7,8,9}, stressing its various 
advantages and improvements. This section just states the results without
derivations.

In section three we will develop the {\bf Master Constraint Programme} for 
a general constrained Quantum Field Theory, the details of General 
Relativity will not be important for that section.

In sections four and five we sketch the construction of a quadratic form 
$Q_{\MC}$ for General Relativity where all the LQG techniques 
developed in \cite{2,3,4,5,6,7,8,9} will be exploited. More precisely,
in section four we describe a graph-changing quadratic form on the 
diffeomorphism invariant Hilbert space which builds directly on the 
techniques of \cite{2,3,4,5,6,7,8,9} while in section five 
we propose an alternative, non-graph-changing quantization of the 
{\bf Master Constraint} which has the advantage of resulting directly 
in a positive, diffeomorphism invariant operator on the kinematical 
Hilbert space ${\cal H}_{Kin}$ (and thus has the Friedrichs extension as 
distinguished self-adjoint extension). This operator can then be induced on
${\cal H}_{Diff}$ and thus sidesteps the quadratic form construction
of section four.
Moreover, it should be possible to verify its classical limit directly with 
the 
methods of \cite{28,29,30,31,32,33,34}. We feel, that this latter
implementation of the {\bf Master Constraint} is less fundamental than 
the one of section four, because it involves an ad hoc quantization step.
Yet, it adds to the faith of the more fundamental quantization since it 
is only a modest modification thereof and can be tested with currently
developed coherent states techniques. In any case it results in a much 
simpler operator and therefore can be easier accessed by analytical
methods.

Finally, in section six we mention directions for further research.
The most important one is presumably that the {\bf Master Constraint
Programme}, at least in th sense of section five, allows for a 
straightforward connection with the spin foam approach of LQG, that is,
a {\bf Path Integral Formulation}. Namely, the {\bf Master Constraint}
can be considered as a {\bf true Hamiltonian}. The difference with 
usual quantum gauge theory is then, besides background independence,
that we are just interested in the kernel of that Hamiltonian. This 
means that our path integral is not a transition amplitude but rather
a generalized projector in the sense of refined algebraic quantization.
Nevertheless, the usual Feynman-Kac like techniques can be employed in 
order to give, hopefully, a rigorous construction of the path integral.\\ 
Another point to mention is that the {\bf Master Constraint Technique} 
could be extended to take care also of the spatial diffeomorphism group.
This is maybe even preferred by those who do not believe in the relevance 
of the spatial diffeomorphism group down to the Planck scale. Moreover,
when doing this we can actually do Hamiltonian (or Lagrangean when
using the path integral) {\bf Lattice LQG} (rather than continuum LQG) 
\cite{34a} without having to be bothered
by the fact that the lattice breaks spatial diffeomorphism invariance.
Spatial diffeomorphism invariance still plays a role, but only on large 
scales, and the {\bf Master Constraint} allows us to take care of this 
new sense of spatial diffeomorphism invariance on any lattice. Finally,
with a new sense of spatial diffeomorphisms, possibly room is made to have 
new kinematical representations of LQG other than the one currently used
\cite{13,14}.

In the appendix we review the notion of Rigging Maps and Rigged Hilbert 
spaces and display the connection with the direct integral method used 
in the main part of this paper. Knowledge of this background material is 
not essential in order to read the paper.

\section{The {\bf Master Constraint Programme} versus the Hamiltonian 
Constraint Quantization Programme for General Relativity} 
\label{s2}

The purpose of this section is to describe the {\bf Master Constraint 
Programme}
when applied to general Relativity. In order to appreciate its advantages
and its technical and conceptual improvements over the original 
Hamiltonian constraint quantization programe, it is 
helpful to recall the quantization of the Hamiltonian constraint 
developed in \cite{2,3,4,5,6,7,8,9} which can be sketched by the following 
steps:
\begin{itemize}
\item[1.] {\it Cotriad Regularization}\\
In order to achieve UV finiteness one had to make sense 
out of a co-triad operator corresponding to to $e_a^j$ where 
$e_a^j e_b^k \delta_{jk}=q_{ab}$ is the spatial metric tensor. 
Since $e_a^j$ is not a polynomial in the elementary phase space 
coordinates $(A_a^j,E^a_j)$ consisting of an $SU(2)$ connection $A$ and a 
canonically conjugate Ad$_{SU(2)}$-covariant vector density $E$ of weight 
one, this could be achieved by writing the co-triad in the form 
$e_a^j=\{A_a^j,V\}$ where $V=\int d^3x \sqrt{\det(q)}$ is the volume 
functional. 
\item[2.] {\it Connection Regularization}\\
Since the connection operator corresponding to $A$ is not 
defined on the Hilbert space one had to write the Poisson bracket in the 
form $\epsilon \{A,V\}=h_{e_\epsilon}(A)\{h_{e_\epsilon}(A)^{-1},V\}$ 
where $h_\epsilon(A)$ denotes the holonomy of $A$ along a path 
$e_\epsilon$ of parameter 
length $\epsilon$. Similarly one had to write the curvature $F$ of $A$
in the form $2\epsilon^2 F=
h_{\alpha_\epsilon}(A)-h_{\alpha_\epsilon}(A)^{-1}$ where 
$\alpha_\epsilon$ denotes a loop of parameter circumference $\epsilon$.
\item[3.] {\it Triangulation}\\
Since the smeared Hamiltonian constraint can be written in the form 
$$
C(N)=\int_\sigma d^3x N \mbox{Tr}(F\wedge e)
$$
where $N$ is a test function, the power $\epsilon^3$ needed in step 2.
can be neatly absorbed by the $d^3x$ volume of a (tetrahedral) cell 
$\Delta$ of a 
triangulation $\tau$ of $\sigma$. Hence one can write the 
Hamiltonian as a 
pointwise (on phase space) limit 
$C(N)=\lim_\epsilon C_\epsilon(N)$ where symbolically
$$
C_\tau(N)=\sum_{\Delta\in \tau} 
\mbox{Tr}([h_{\alpha_\Delta}(A)-h_{\alpha_\Delta}(A)^{-1}]
h_{e_\Delta}(A)\{h_{e_\Delta}(A)^{-1},V\})
$$ 
\item[4.] {\it Quantization of the Regulated Constraint}\\
In order to quantize this expression one now replaced all appearing 
quantities by operators and the Poisson bracket by a commutator divided by 
$i\hbar$. In addition, in order to arrive at an unambiguous result one had 
to make the triangulation {\it state dependent}. That is, the 
regulated operator 
is defined on a certain (so-called spin network) basis elements 
$T_s$ of the 
Hilbert space in terms of an adapted triangulation $\tau_s$ and extended 
by linearity. This is justified because the 
Riemann sum that enters the definition of $C_\tau(N)$ converges 
to $C(N)$ no matter how we refine the triangulation. 
%
\item[5.] {\it Removal of the Regulator}\\
To take the infinite refinement limit (or continuum) $\tau\to \sigma$ 
of the resulting regulated operator 
$\hat{C}^\dagger_\tau(N)$ is non-trivial because the holonomy 
operators 
$\hat{h}_{e_\Delta}$ are not even weakly continuously represented on the 
Hilbert space, hence the limit cannot exist in the weak operator topology.
It turns out that it exists in the, what one could call, {\it weak
Diff$^\ast$ topology} \cite{1}: Let ${\Phi}_{Kin}$ be a dense invariant 
domain for 
the (closable) operator $\hat{C}^\dagger_\tau(N)$ on the Hilbert space 
${\cal H}_{Kin}$ and let $(\Phi_{Kin}^\ast)_{Diff}$ be 
the set of all spatially diffeomorphism invariant distributions over 
$\Phi_{Kin}$
(equipped with the topology of pointwise convergence). Then 
$\lim_{\tau\to \sigma} \hat{C}_\tau(N)=\hat{C}(N)$ if and only if
for each $\epsilon>0,\;T_s,\;l \in(\Phi^\ast_{Kin})_{Diff}$ 
there exists 
$\tau_s(\epsilon)$ {\it independent of $l$} such that 
$$
|l([\hat{C}_{\tau_s}(N)-\hat{C}(N)]T_s)|<\epsilon\;\;\forall\;\;
\tau_s(\epsilon)\subset \tau_s 
$$
That the limit is uniform in $l$ is crucial because it excludes
the existence of the limit on spaces larger than 
$(\Phi_{Kin}^\ast)_{Diff}$
\cite{16,17} which would be unphysical because the space of solutions 
to all constraints must obviously be a subset of 
$(\Phi_{Kin}^\ast)_{Diff}$. Notice that the limit is required refinements
of adapted triangulations only.

What we just said is sometimes paraphrased by saying that ``the 
Hamiltonian constraint is defined only on diffeomorphism invariant 
states". But this is certainly wrong in the strict sense, the Hamiltonian 
constraint is not diffeomorphism invariant and thus the dual operator
defined by $[\hat{C}(N)'\Psi](f):=\Psi(\hat{C}(N)^\dagger f)$ does not 
preserve 
$(\Phi_{Kin}^\ast)_{Diff}$. Rather, in a technically precise sense the 
constraint
is defined on the Hilbert space ${\cal H}_{Kin}$ itself. In order to write 
it down 
explicitly, one needs to make use of the axiom of choice,
$\hat{C}(N) T_s:=\hat{C}_{\tau^0_s}(N) T_s$ in the sense that
there is an explicit action which is defined up to a diffeomorphism and 
the choice $s\mapsto \tau^0_s$ involved corresponds to the choice of a 
diffeomorphism. One 
does not need to worry about this choice since it is irrelevant on the 
space of full solutions which is in particular a subset of 
$(\Phi_{Kin}^\ast)_{Diff}$.
\item[6.] {\it Quantum Dirac Algebra}\\
We may now compute the commutator $[\hat{C}(N),\hat{C}(N')]$ on 
$\Phi_{Kin}$
corresponding to the Poisson bracket $\{C(N),C(N')\}$ which is 
proportional to the spatial diffeomorphism constraint $C_a$.
This commutator turns out to be non-vanishing on $\Phi_{Kin}$ as it should 
be,
however, $\Psi([\hat{C}(N),\hat{C}(N')]f)=0$ for all $f\in \Phi_{Kin},
\Psi\in (\Phi_{Kin}^\ast)_{Diff}$. This is precisely how we would expect 
it 
in the absence of an anomaly. Note that this is sometimes paraphrased 
by ``The algebra of Hamiltonian Constraints is Abelean". But this is clearly
wrong in the strict sense, the commutator is defined on $\Phi_{Kin}$, 
where it does not vanish, and not on $(\Phi_{Kin}^\ast)_{Diff}$. On the 
other hand, the right hand side of 
the commutator on $\Phi_{Kin}$ does not obviously resemble the 
quantization 
of the classical expression $\int d^3x (N N'_{,a}- N_{,a} N') q^{ab} C_b$
so there are doubts, expressed in \cite{15,16,17} whether the 
quantization of $C(N)$ produces the correct quantum dynamics. Notice, 
however, that this is not surprising because in order to write 
the classical Poisson bracket $\{C(N),C(N')\}$ in the form 
$\int d^3x (N N'_{,a}- N_{,a} N') q^{ab} C_b$ one must perform 
integrations by parts, use non-linear differential geometric identities,
reorder terms etc., which are manipulations all of which are very 
difficult to perform at the quantum level. Moreover, it is possible to
quantize the expression $\int d^3x (N N'_{,a}- N_{,a} N') q^{ab} V_b$ 
directly \cite{5} and its dual annihilates $(\Phi_{Kin}^\ast)_{Diff}$ as 
well. 

In summary: There is certainly no mathematical inconsistency but there 
are doubts on the correct classical limit of the theory. Notice, 
however, that aspects of the Hamiltonian constraint quantization has been 
successfully tested in model systems \cite{4,18}.
\item[7.] {\it Classical Limit}\\
In order to improve on this one could try to prove the correctness of the 
Hamiltonian constraint by computing its expectation value in coherent 
states for non-Abelean gauge theories \cite{28,29,30,31,32,33} because 
then the manipulations just mentioned can be performed for the 
function valued (rather than operator valued) expectation values. However,
the problem with such an approach is that coherent states which would 
suitable for doing this would naturally be elements of the full 
distributional dual $\Phi_{Kin}^\ast$ of $\Phi_{Kin}$ as has been shown in 
\cite{34,35}.
However, $\Phi_{Kin}^\ast$, in contrast to $(\Phi_{Kin})^\ast_{Diff}$, does 
not carry 
a (natural) inner product, so that expectation values cannot be computed. 
The 
technical reason for why the non-distributional states constructed in 
\cite{28,29,30,31,32,33,34} are insufficient to compute the classical 
limit is that they are designed 1) for a representation which supports 
the non-Abelean holonomy-flux algebra \cite{13,14} and 2) for 
operators only which leave invariant 
the graph on which a given spin network state is supported, however, the 
Hamiltonian constraint does not naturally have this latter property.
\end{itemize}
The {\bf Master Constraint Approach} now improves on these issues as follows:
\begin{itemize}
\item[1.+2.+3.] {\it Cotriad Regularization, Connection Regularization 
and Triangulation}\\ 
These three steps are essentially unchanged.
\item[4.] {\it Quantization of the Regulated Constraint}\\
It is here where things become most interesting: In contrast to 
$\hat{C}(N)$
which had to be quantized on $\Phi_{Kin}$ since $\hat{C}(N)$ does not 
preserve
diffeomorphism invariant states, we can quantize $\MC$ directly on 
$(\Phi_{Kin}^\ast)_{Diff}$, {\it thus being able to use the full power of 
the results
derived in \cite{26}}. We can sketch the procedure as follows: There is an
anti-linear map $\eta:\;\Phi_{Kin}\to (\Phi_{Kin}^\ast)_{Diff}$ 
constructed in 
\cite{26} the image of which is dense in the Hilbert space 
${\cal H}_{Diff}$ of diffeomorphism invariant states which carries the inner 
product $<\eta(f),\eta(f')>_{Diff}:=\eta(f')[f]$ where the right hand 
side denotes the action of the distribution $\eta(f')$ on the element 
$f\in \Phi_{Kin}$. We now point-split $\MC$ as follows
$$
\MC=\lim_{\tau\to\sigma \atop \epsilon\to 0} \sum_{\Delta,\Delta'}
[\int_\Delta d^3x (\frac{C}{\sqrt[4]{\det(q)}})(x)]
[\int_{\Delta'} d^3y \phi_\epsilon(x,y) (\frac{C}{\sqrt[4]{\det(q)}})(y)]
$$
where $\phi_\epsilon(x,y)$ is any point splitting function converging 
to the $\delta-$distribution in the sense of tempered distributions
and $\Delta,\Delta'$ are cells of the triangulation.
Exploiting that $<.,.>_{Diff}$ has a complete orthonormal basis 
$\eta(b_I)$ for some $b_I\in \Phi_{Kin}$ where $I$ belongs to some index 
set 
$\cal I$, we try to define the {\it quadratic form} on (a suitable form 
domain of) ${\cal H}_{Diff}$ by  
$$
Q_{\MC}(\eta(f),\eta(f'))
:=\lim_{\tau\to\sigma \atop \epsilon\to 0} \sum_{I\in {\cal I}}
\int_\sigma\; d^3x\; \int_\sigma\; d^3y\; \phi_\epsilon(x,y) \;
\overline{\eta(f)[\widehat{\frac{C}{\root 4\of{\det(q)}}}(x) b_I]}\;
\eta(f')[\widehat{\frac{C}{\root 4\of{\det(q)}}}(y) b_I]
$$
Now it turns out that because we were careful enough to keep the 
integrand of $\MC$ a density of weight one, the limit $\epsilon\to 0$ 
can be taken with essentially the same methods as those developed in 
\cite{2,3,4,5,6}. The operator valued distributions that appear in that 
expression involve a loop derivative, however, since we are working with 
diffeomorphism invariant states, {\it we never have to take the derivative,
we just need to take loop differentials}! This is in contrast to \cite{37}
and again related to the density weight one. 
\item[5.] {\it Removal of the Regulator}\\
Now by the same methods as in \cite{2,3,4} one can remove the 
triangulation dependence. Diffeomorphism invariance ensures that the limit
does not depend on the representative index set $\cal I$ (the map $\eta$
is many to one).

In summary, we end up with a quadratic form $Q_{\MC}$ on 
${\cal H}_{Diff}$ which by inspection is {\it 
positive}, hence semibounded.
Now, if we can prove that the form is closed, then there is a unique 
positive, self-adjoint operator $\MCO$ such that
$$
Q_{\MC}(\eta(f),\eta(f'))=
<\eta(f),\MCO \eta(f')>_{Diff}
$$
Notice that this operator is automatically densely defined and closed on 
${\cal H}_{Diff}$, so we really have pushed the constraint analysis one 
level up
from ${\cal H}_{Kin}$. Notice that not every semibounded quadratic form is 
closable
(in contrast to symmetric operators) and therefore whether or not 
$Q_{\MC}$ is closable is a non-trivial and crucial open question 
which we are going to address in a subsequent publication \cite{42a}.
\item[6.] {\it Quantum Dirac Algebra}\\
There is no constraint algebra any more, the issue of mathematical 
consistency (anomaly freeness) is trivialized. However, the issue of 
physical consistency is not answered yet in the sense that operator
ordering choices will have influence on the size of the physical Hilbert 
space and thus on the number of semiclassical states, see below.
\item[7.] {\it Classical Limit}\\
Since no semiclassical states have been constructed yet on ${\cal H}_{Diff}$
we cannot decide whether the {\bf Master Constraint Operator}
$\MCO$, if it exists, has the correct classical limit. 
However, the fact that there is significantly less quantization ambiguity 
involved than for the Hamiltonian constraints $C(x)$ themselves adds some
faith to it. Moreover, the issues of 7. above 
could improve on the level of ${\cal H}_{Diff}$ for two reasons:
First of all, ${\cal H}_{Diff}$ in contrast to $\Phi_{Kin}^\ast$ {\it does} 
carry an inner product. Secondly, when adopting
the viewpoint of section \ref{s4.4}, the Hilbert space ${\cal H}_{Diff}$
is separable and hence coherent states are not distributional but rather
honest elements of ${\cal H}_{Diff}$. Finally, there is a less ambitious 
programme which we outline in section \ref{s5} where $\MCO$ exists
as a diffeomorphism invariant operator on ${\cal H}_{Kin}$ and where one 
can indeed
try to answer the question about the correctness of the classical limit 
with existing semiclassical tools. The reason for why this procedure is less 
favoured is that
it is an ad hoc modification of the action of the operator in such a way that
it leaves the graph of a spin network on which the operator acts invariant.
On the other hand, the modification is not very drastic and 
therefore supports our more fundamental version of the {\bf Master 
Constraint Operator}.
\end{itemize}
In addition to these technical improvements of \cite{2,3,4,5,6,7}, the {\bf 
Master 
Constraint Operator}, 1. if it exists and 2. if the issue about its 
classical limit can be settled, also possibly provides conceptual 
advantages because {\it it has a 
chance to complete the technical steps of the Canonical 
Quantization Programme:} 
\begin{itemize}
\item[i.] {\it Solution of all Quantum Constraints}\\
Since by construction $\MCO$ is a self-adjoint 
operator we can solve the {\bf Master Constraint} in the following,
at least conceptually simple, way which rests on the assumption that 
${\cal H}_{Diff}$ {\it is 
separable}, that is, the index set $\cal I$ has countable cardinality.
Now as is well known \cite{38} the Hilbert space is not a priori separable
because there are continuous moduli associated with intersecting knot 
classes with vertices of valence higher than four. It turns out that 
there is a simple way to remove those moduli by performing an additional 
averaging in the definition of the (rigging) map $\eta$ mentioned above.
This should not affect the classical limit of the theory because this 
modification is immaterial for vertices of valence four or lower which 
are the ones that are most important in semiclassical considerations of 
lattice gauge theories (essentially because they are sufficient to 
construct dual triangulations). If one is satisfied with heuristic 
considerations then we can use Rigging Map techniques to solve the 
constraint which formally work also in the non-separable case, see the 
appendix.

Thus, if ${\cal H}_{Diff}$ is separable, then we can construct the direct
integral representation of ${\cal H}_{Diff}$ associated with the 
self-adjoint operator $\MCO$, that is,
$$
{\cal H}_{Diff}=\int_{\Rl}^\oplus d\mu(\lambda) {\cal 
H}^\oplus_{Diff}(\lambda) $$
where $\mu$ is a positive probability measure on $\mu$.
Notice that the scalar product on the individual 
${\cal H}^\oplus_{Diff}(\lambda)$ and the measure $\mu$ are uniquely 
induced by that on ${\cal H}_{Diff}$ up to unitary equivalence.
Since  
$\MCO$ acts on the Hilbert space
${\cal H}^\oplus_{Diff}(\lambda)$ by multiplication with $\lambda$ it 
follows that
$$
{\cal H}_{Phys}:={\cal H}^\oplus(0)
$$
is the physical Hilbert space and a crucial open question to be answered is 
whether it is large enough (has a sufficient number of 
semiclassical solutions). 
\item[ii.] {\it Quantum Dirac Observables}\\
Let $\hat{O}_{Diff}$ be a bounded self adjoint operator on ${\cal 
H}_{Diff}$, for 
instance 
a spectral projection of the volume operator corresponding to the total 
volume of $\sigma$. Since the {\bf Master Constraint Operator} is 
self-adjoint, we may construct the strongly continuous one-parameter 
family of unitarities 
$\hat{U}(t)=\exp(it\MCO)$. Then, if the uniform limit 
exists, the operator 
$$
[\hat{O}_{Diff}]:=\lim_{T\to \infty} \frac{1}{2T} \int_{-T}^T dt\;
\hat{U(t)}\; \hat{O}_{Diff} \;\hat{U}(t)^{-1}
$$
commutes with the {\bf Master Constraint}, hence provides a strong Dirac 
observable, preserves ${\cal H}_{Phys}$ and induces a bounded self-adjoint 
operator $\hat{O}_{Phys}$ there.
\end{itemize}
Should all of these steps go through we would still be left with conceptual
issues such as interpretational ones, the reconstruction problem etc.,
see e.g. \cite{39}, as well as practical ones (actually computing things). 
However, at least on the technical side (i.e. rigorous existence) the 
steps outlined above look very promising because they involve standard 
functional analytic questions which are well posed and should have 
a definite answer. They are much cleaner than the ones involved in the 
Hamiltonian Constraint Quantization programme. Let us list those questions 
once more: 
\begin{itemize}
\item[Task A] {\it Closure of Quadratic Form}\\
Show that the quadratic form $Q_{\MC}$ is closable. This is 
an important task because it is not granted by abstract theorems
that this is possible. Once the existence of the closure is established
then it is only a practical problem to 
compute the unique positive, self-adjoint operator 
$\MCO$ to which it corresponds. 
\item[Task B] {\it Spectral Analysis of $\MCO$}\\
Derive a direct integral representation of the Hilbert space 
${\cal H}_{Diff}$ corresponding to $\MCO$.
Also this is only a practical problem once we have made ${\cal H}_{Diff}$
separable by the additional averaging described below. The 
corresponding physical Hilbert space is then simply 
${\cal H}_{Phys}={\cal H}_{Diff}^\oplus(0)$. Show that it is ``large 
enough",
that is, contains a sufficient number of physical, semiclassical states. 
\item[Task C] {\it Quantum Dirac Observables}\\
Find out for which diffeomorphism invariant, bounded, self-adjoint 
operators 
$\hat{O}_{Diff}$ the corresponding ergodic mean $[\hat{O}_{Diff}]$
converges (in the uniform operator topology induced by the topology of 
${\cal H}_{Diff}$). Then compute the induced operator $\hat{O}_{Phys}$ on
${\cal H}_{Phys}$ which is automatically self adjoint.
\item[Task D] {\it Spatially Diffeomorphism Invariant Coherent States}\\
Construct semiclassical, spatially diffeomorphism invariant states, maybe
by applying the map $\eta$ to the states constructed in  
\cite{28,29,30,31,32,33,34}, and compute expectation values and 
fluctuations of the {\bf Master Constraint Operator}. Show that these 
quantities coincide with the expectetd classical values up to $\hbar$ 
corrections. This is the second most important step because the existence 
of suitable semiclassical states, at the spatially diffeomorphism 
invariant level is not a priori granted. Once this step is 
established, we would have shown that the classical limit of 
$\MCO$ is the correct one and therefore the quantization
really qualifies as quantum field theory of GR. Notice that this task 
can be carried out even before we show that $Q_{\MC}$ is closable. 
\end{itemize}

\section{Elements of the {\bf Master Constraint Programme}}
\label{s3}

In this section we will describe in more detail the basic ideas of the 
{\bf Master Constraint Programme} for a general theory. Presumably 
elements of that idea are scattered over the vast literature on
constraint quantization but to the best knowledge of the author these 
elements have not been combined into the form that we need here which is 
why we included this section. Also we believe that one can state the 
results outlined below in much broader contexts and it is very well 
possible that hard theorems already exist in the literature that state them 
more precisely than we do here. The author would be very interested to 
learn about the existence of such results.

\subsection{Classical Theory}
\label{s3.1}

We begin with a general symplectic manifold $({\cal M},\;\{.,.\})$ which 
may be infinite dimensional. Here $\cal M$ is a differentiable manifold 
modelled on a Banach space (phase space) and the (strong) symplectic 
structure is defined in terms of a Poisson bracket $\{.,.\}$ on
$C^\infty({\cal M})\times C^\infty({\cal M})$. Furthermore, we are given 
a set of constraint functions $C_J,\;J\in {\cal J}$ where the index set 
$\cal J$ may involve discrete and continuous labels. For the sake of 
definiteness and because it is the most interesting case in field theory, 
suppose that ${\cal J}=D\times X$ where $D$ is a discrete label set 
and $X$ is a topological space. Hence we may write $J=(j,x)$ and 
$C_J=C_j(x)$. Without loss of generality we assume that all constraints 
are real-valued and first class, replace the Poisson bracket by the 
corresponding Dirac bracket in order to remove potentially present
second class constraints if necessary.

Let $\mu$ be a positive measure on the Borel $\sigma-$algebra of $X$ and 
for each $x\in X;\;j,k\in D$ let be given a positive definite ``metric" 
function $g^{jk}(x)\in C^\infty({\cal M})$. We will now state some 
more or less obvious results which we, however, could not find anywhere 
in the literature.
\begin{Lemma} \label{la3.1} ~~~\\
The constraint hypersurface $\cal C$ of $\cal M$ defined by 
\be \label{3.1}
{\cal C}:=\{m\in{\cal M};\;\; C_J(m)=0\;\mbox{ for } \mu-\mbox{a.a. } 
J\in {\cal J}\} 
\ee
can be equivalently defined by
\be \label{3.2}
{\cal C}=\{m\in {\cal M};\;\; \MC(m)=0\} 
\mbox{ ~~~{\bf (Master Condition)}}
\ee
where 
\be \label{3.3}
\MC:=\frac{1}{2}\int_X\; d\mu(x) \sum_{j,k\in D} q^{jk}(x) C_j(x) C_k(x)
\ee
is the called {\bf Master Constraint} associated with $C_J,\mu,g$.
\end{Lemma}
The proof is trivial. If, as usually the case in applications, the 
functions $C_J(m), g(m)$ are smooth in $X$ for each $m\in {\cal M}$ then 
the ``almost all" restriction can be neglected.

Let us recall the notion of a Dirac observable.
\begin{Definition} \label{def3.1} ~~~\\
i)\\
A function $O\in C^\infty({\cal M})$ is called a weak Dirac observable 
provided that 
\be \label{3.4}
\{O,C(N)\}_{|{\cal C}}=0
\ee
for all test functions (smooth of compact support) $N^j$ and where 
\be \label{3.5}
C(N)=\int_X\; d\mu(x) \; N^j(x) C_j(x)
\ee
The set of all weak Dirac observables is denoted by ${\cal O}_w$.\\
ii)\\
A function $O\in C^\infty({\cal M})$ is called an ultrastrong Dirac 
observable provided that 
\be \label{3.6}
\{O,C(N)\}\equiv 0
\ee
for all test functions (smooth of compact support) $N^j$.
The set of all ultrastrong Dirac observables is denoted by ${\cal O}_u$.
\end{Definition}
Obviously every ultrastrong Dirac observable is a weak Dirac observable.
The following simple theorem is crucial for the validity of the {\bf Master 
Constraint Proposal}. 
\begin{Theorem} \label{th3.1} ~~~\\
A function $O\in C^\infty({\cal M})$ is a weak Dirac Observable if and 
only if
\be \label{3.7}
\{O,\{O,\MC\}\}_{\MC=0}=0
\mbox{ ~~~{\bf (Master Equation)}}
\ee
\end{Theorem}
Proof of theorem \ref{th3.1}:\\
The proof is so trivial that we almost do not dare to call this result a 
theorem:

Since $O$ is certainly twice differentiable by assumption we easily 
compute (formally, to be made precise using the topology of $\cal M$)
\ba \label{3.8}
\{O,\{O,\MC\}\} &=& \int_X d\mu(x)\; 
 [g^{jk}(x)\{O,C_j(x)\}\{O,C_k(x)\}+
  g^{jk}(x)\{O,\{\{O,C_j(x)\}\}C_k(x)
\nonumber\\
&& + \{O,g^{jk}(x)\}\{O,C_j(x)\}C_k(x)+
  \frac{1}{2} \{O,\{O,g^{jk}(x)\}\}C_j(x) C_k(x)]
\ea
Restricting this expression to the constraint surface $\cal C$ is 
equivalent, according to lemma \ref{la3.1}, to setting $\MC=0$ hence 
\be \label{3.9}
\{O,\{O,\MC\}\}_{\MC=0}=\int_X d\mu(x)\; 
 g^{jk}(x)\{O,C_j(x)\}_{|{\cal C}}\{O,C_k(x)\}_{|{\cal C}}
\ee
Since $g$ is positive definite this is equivalent with 
\be \label{3.10}
\{O,C_j(x)\}_{|{\cal C}}=0 \mbox{ for a.a. }x\in X
\ee
hence this is equivalent with
\be \label{3.11}
\{O,C(N)\}_{|{\cal C}}=0
\ee
for all smooth test functions of compact support.\\
$\Box$\\
Obviously the theorem also holds under the weaker assumption that 
$O$ is at least twice differentiable.

The characterization of weak Dirac observables as shown in theorem 
\ref{th3.1} motivates the following definition.
\begin{Definition} \label{def3.2} ~~~\\
A function $O\in C^\infty({\cal M})$ is called strong Dirac 
Observable if 
\be \label{3.12}
\{O,\MC\} \equiv 0
\ee
The set of all strong Dirac observables is denoted by ${\cal O}_s$.
\end{Definition}
The inclusion ${\cal O}_s\subset{\cal O}_w$ is obvious. Notice that
restriction to twice differentiable functions does not harm the validity 
of this inclusion. However, whether
every ultrastrong Dirac observable is also a strong one is less clear as 
one can easily check and presumably there do exist counterexamples. We
will not be concerned with this in what follows.
\begin{Definition} \label{def3.3} ~~~\\
Let $O\in C^\infty_b({\cal M})$ be a bounded function on $\cal M$
(in sup-norm) and let 
$\chi_{\MC}$
be the Hamiltonian vector field of $\MC$ (which is uniquely determined  
because our symplectic structure is strong by assumption). The 
{\bf ergodic mean} of $O$, if it exists, is defined by the pointwise
(on $\cal M$) limit
\be \label{3.13}
[O]:=\boldsymbol{\mathsf{P}}^{\MC}\cdot O:=\lim_{T\to\infty} \frac{1}{2T} 
\int_{-T}^T \;dt \; e^{t{\cal L}_{\chi_{\MC}}} O
\ee
\end{Definition}
It is clear that the operator 
$\boldsymbol{\mathsf{P}}^{\MC}:\;
C_b({\cal M})\to C_b({\cal M})$ is (formally) a projection 
because the operators 
\be \label{3.14}
\alpha^{\MC}_t:=e^{t{\cal L}_{\chi_{\MC}}}
\ee
are unitary on the Hilbert space $L_2({\cal M},d\Omega)$ where 
$\Omega$ is the (formal) Liouville measure on $\cal M$. Notice that 
the one-parameter family of symplectomorphisms (\ref{3.14}) defines 
an inner automorphism on the Poisson algebra $C^\infty({\cal M})$.

The usefulness of the notion of a strong Dirac observable is stressed by 
the following result.
\begin{Theorem} \label{th3.2} ~~~\\
Suppose that $[O]$ is still at least in $C^2_b({\cal M})$. 
Then, under the assumptions spelled out in the proof, $[O]\in {\cal O}_s$.
\end{Theorem}
Notice that the requirement that $[O]$ is at least twice differentiable 
is crucial, otherwise $[O]$ is not granted to be an element of ${\cal O}_w$ 
even if it has vanishing first Poisson bracket with $\MC$ everywhere 
on $\cal M$.\\
Proof of theorem \ref{th3.2}:\\
Let 
\be \label{3.15}
[O]_T:=
\frac{1}{2T} \int_{-T}^T \;dt \; e^{t{\cal L}_{\chi_{\MC}}} O
\ee
Provided we may interchange the integral with the Poisson bracket we 
have 
\be \label{3.16}
\{[O]_T,\MC\}=
\frac{1}{2T} \int_{-T}^T \;dt \; \frac{d}{dt} e^{t{\cal L}_{\chi_{\MC}}} O
=\frac{e^{T{\cal L}_{\chi_{\MC}}}-e^{-T{\cal L}_{\chi_{\MC}}}}{2T} O
\ee
Since $O$ is bounded (in sup-norm) on $\cal M$ by assumption, so is 
$e^{\pm T{\cal L}_{\chi_{\MC}}} O$, hence 
\be \label{3.17}
\lim_{T\to \infty} \{[O]_T,\MC\}=0
\ee
Thus, provided that we may interchange the limit $T\to \infty$ with the 
Poisson bracket, we get $\{[O],\MC\}=0$.\\
$\Box$\\
Sufficient conditions for the existence assumptions and the allowedness 
to interchange the opertions indicated in the proof are examined in 
\cite{41}.
The restriction of the ergodic mean to bounded functions is thus 
motivated by the proof of this theorem. On the other hand, in order that
$[O]\not=0$ it is necessary that the evolution $\alpha^{\MC}_t(O)$ does not
decay (in sup-norm) as $t\to\infty$ but rather stays bounded away from zero.

Hence the ergodic mean technique provides a {\it guideline for explicitly
constructing Dirac Observables}. It would be nice to have an equally 
powerful technique at our disposal which immediately constructs 
functions which satisfy the more general non-linear condition (\ref{3.7}) 
but we could not find one yet. Notice that, formally, we can carry out 
the integral in (\ref{3.13}) by using the power expansion, valid for 
smooth $O$
\be \label{3.18}
\alpha^{\MC}_t(O)=\sum_{n=0}^\infty \;\frac{t^n}{n!} \;\{O,\MC\}_{(n)}
\ee
where the the multiple Poisson bracket is inductively defined by 
$\{O,\MC\}_{(0)}=O,\;\{O,\MC\}_{(n+1)}=\{\{O,\MC\}_{(n)},\MC\}$.
Hence, formally 
\be \label{3.19}
[O]_T=\sum_{n=0}^\infty \;\frac{T^{2n}}{(2n+1)!} \;\{O,\MC\}_{(2n)}
\ee
However, the series is at best asymptotic and the limit $T\to \infty$ 
can be taken only after resummation of the series. Thus the presentation
(\ref{3.19}) is useless, we need a perturbative definition 
of $[O]$ which is practically more useful with more control on 
convergence issues. Maybe some of the ideas of \cite{39a} can be 
exploited. This leads us to the {\bf theory
of dynamical systems, integrable systems, the theory of invariants and 
ergodic theory} \cite{40}. Aspects of this are currently under 
investigation and will be published elsewhere \cite{41}.  \\
\\
A couple of remarks are in order before we turn to the quantum theory:\\
1.\\
When applying the above theory to LQG at the diffeomorphism invariant 
level
one should start the classical description from the diffeomorphism 
invariant phase space. This is defined by considering
all diffeomorphism invariant functions on the kinematical phase space 
as its (over)coordinatization and by using the Poisson bracket between 
diffeomorphism invariant functions induced by the kinematical Poisson bracket
as the diffeomorphism invariant Poisson bracket between those functions.\\
2.\\
The automorphisms $\alpha^{\MC}_t$ do not only preserve the constraint 
hypersurface $\cal C$, they also preserve {\it every individual point}
$m\in {\cal C}$ because $\{O,\MC\}_{\MC=0}=0$ for any 
$O\in C^\infty({\cal M})$. In other words, the Hamiltonian vector field
$\chi_{\MC}$ is not only tangential to $\cal C$ as it is the case for 
first class constraints, it vanishes identically on $\cal C$. Imagine 
a foliation of $\cal M$ by leaves ${\cal M}_t=\{m\in {\cal M};\;
\MC(m)=t\}$ where ${\cal M}_0={\cal C}$. If $\MC$ is an least once 
differentiable function on $\cal M$ determined by first class 
constraints as we assumed then in an open neighbourhood of $\cal C$ the 
vector field $\chi_{\MC}$ will be tangential to the corresponding leaf 
but non-vanishing there. This is 
dangerous because it means that the 
automorphisms $\alpha^{\MC}_t$ for $t\not=0$
are non-trivial and thus the ergodic mean $[O]$ could be discontinuous 
precisely in any neighbourhood of $\cal C$, hence it is not 
differentiable there and thus does not qualify as a strong Dirac observable.
Investigations in simple models show that this indeed happens. 
In fact, experience with dynamical systems reveals that 
typical observables (integrals of motion generated by the ``Hamiltonian" 
$\MC$) generated by an ergodic mean are rather discontinuous functions on 
$\cal M$ even if the system is completely integrable \cite{42}. 

However, there
is a simple procedure to repair this: 
Consider only the values of $[O]$ restricted to a set 
of the form ${\cal M}-{\cal U}$ where $\cal U$ is any open neighbourhood 
of $\cal C$. If $[O]$ is at least $C^2$ there, extend to all of 
$\cal M$ in an at least $C^2$ fashion. The resulting new observable
$[O]'$ coincides with $[O]$ except on $\cal C$, is $C^2$ and satisfies
$\{[O]',\MC\}\equiv 0$ hence defines an element of ${\cal O}_w$.\\ 
3.\\
One could avoid these subtleties by generalizing the concept of a strong
Dirac observable as follows:
\begin{Definition} \label{def3.4} ~~~\\
Let $\alpha^{\MC}_t$ be the one-parameter family of automorphisms 
(symplectic isometries) generated
by the Hamiltonian vector field $\chi_{\MC}$ of the {\bf Master Constraint}.
Then a (not necessarily continuous) function on $\cal M$ is called a 
generalized strong Dirac observable provided that
\be \label{3.20}
\alpha^{\MC}_t(O)=O
\ee
for all $t\in \Rl$.
\end{Definition}
Notice that definitions \ref{def3.4} and \ref{def3.2} are certainly not 
eqivalent. On the other hand,
with this weaker notion the above discontinuity problems disappear and 
the proof of theorem \ref{th3.2} holds under weaker assumptions because 
now limits and integrals commute under weaker assumptions with the operation 
$\alpha^{\MC}_t$. Also in quantum theory definition \ref{def3.4} is 
easier to deal with because it changes the focus from the unbounded, 
self-adjoint operator $\MC$ to the bounded, unitary operators 
$\exp(it \MCO)$ which avoids domain questions, see below. \\
4.\\
In summary, the ergodic mean $[O]$ of $O$ is a {\bf candidate} for an element
of ${\cal O}_s$. In order to check whether it really is it is sufficient 
that it be twice differentiable and $\{O,\MC\}\equiv 0$. 
Similarly, in the quantum theory one must check whether the 
{\bf candidate} operator $[\hat{O}]$ of (\ref{3.24}) has commuting 
spectral projections with those of $\MCO$ since in general only then 
$[[\hat{O}],\MCO]\equiv 0$ identically. There is no differentiability
condition to be checked, however, since the quantum analogue of
Poisson brackets, namely commutators between bounded spectral projections,
always make sense. In that sense the quantum theory is better behaved.

\subsection{Quantum Theory}
\label{s3.2}

Let us now come to quantization which is actually just a straightforward 
transcription of the above structure from functions to operators. 
Suppose that we 
managed to find a representation of the Poisson algebra as an algebra 
$\widehat{{\cal O}}$ of operators on a {\it 
separable} Hilbert 
space ${\cal H}_{Kin}$ and that, in particular, the {\bf Master 
Constraint} $\MC$
is represented as a {\it positive, self-adjoint} operator $\MCO$. We 
therefore can construct the weakly continuous, one-parameter family of
unitary operators 
\be \label{3.21}
\hat{U}(t)=e^{it\MCO}
\ee
We then have a representation of the automorphisms $\alpha^{\MC}_t$ of 
$\cal O$ as inner automorphisms on $\widehat{{\cal O}}$ according to
\be \label{3.22}
\hat{\alpha}^{\MC}_t(\hat{O})=\hat{U}(t)\hat{O}\hat{U}(t)^{-1}
\ee
\begin{Definition} \label{def3.5}
A strong, {\bf generalized} quantum Dirac observable 
is defined as a self-adjoint element $\hat{O}\in \widehat{{\cal O}}$ 
such that 
\be \label{3.23}
\hat{\alpha}^{\MC}_t(\hat{O})=\hat{O}
\ee
for all $t\in \Rl$. A (genuine) strong quantum Dirac Observable is such 
that the spectral projections of $\hat{O}$ and $\MCO$ commute.
\end{Definition}
Candidates of the set ${\cal O}_s$ of strong quantum Dirac Observables 
can be constructed using the ergodic mean of bounded operators
$\hat{O}$ (in the uniform operator norm 
induced by the topology of ${\cal H}_{Kin}$)
\be \label{3.24}
[\hat{O}]=\lim_{T\to\infty} \frac{1}{2T} \int_{-T}^T\;dt\; 
\hat{\alpha}^{\MC}_t(\hat{O})
\ee
Notice that no domain questions arise. The direct translation of 
(\ref{3.7}) is more difficult since it cannot be easily written in terms 
of the bounded operators $\hat{U}(t)$, hence domain questions arise:
\begin{Definition} \label{def3.6}
Let $\Phi_{Kin}$ be a common, dense, invariant domain for the self adjoint
operators $\hat{O},\MCO$ and let $(\Phi_{Kin}^\ast)_{Phys}$ be the subspace 
of the space $\Phi_{Kin}^\ast$ of algebraic distributions over $\Phi_{Kin}$
satisfying $l(\MCO f)=0$ for any $f\in \Phi_{Kin}$. Then $\hat{O}$
is a weak quantum Dirac Observable provided that 
\be \label{3.24a}
l([\hat{O},[\hat{O},\MCO]]f)=0 \mbox{ for all }
f\in\Phi_{Kin},\; l\in(\Phi_{Kin}^\ast)_{Phys}
\ee
\end{Definition}
Obviously any strong quantum Dirac observable is a weak one. 
A more precise examination between the notions of strong and weak quantum 
Dirac observables will be performed in \cite{41}. \\
\\
As shown in the appendix, there is a one to one correspondence between 
$l_\psi\in (\Phi_{Kin}^\ast)_{Phys}$ (defined below) and solutions 
to the quantum constraint, i.e physical states $\psi\in 
{\cal H}_{Phys}$, heuristically given by $l_\psi=<\delta(\MCO)\psi,.>_{Kin}$.
The precise construction of physical states is conceptually straightforward:
Since ${\cal H}_{Kin}$ is separable by assumption, as is well-known, 
\cite{43} it can be 
represented as a direct integral of separable Hilbert spaces 
${\cal H}^\oplus_{Kin}(\lambda),\;\lambda\in \Rl$, subordinate to $\MCO$ 
according to 
\be \label{3.25}
{\cal H}_{Kin}=\int^\oplus_{\Rl} \;d\nu(\lambda) \; 
{\cal H}^\oplus_{Kin}(\lambda)
\ee
that is, any element $\psi\in {\cal H}_{Kin}$ can be thought of as a 
collection
$(\psi(\lambda))_{\lambda\in \Rl}$ where 
$\psi(\lambda)\in {\cal H}^\oplus_{Kin}(\lambda)$ such that 
\be \label{3.26}
||\psi||^2=\int_\Rl\; d\nu(\lambda)\;
||\psi(\lambda)||^2_{{\cal H}^\oplus_{Kin}(\lambda)}
\ee
converges (the functions $\lambda\mapsto
||\psi(\lambda)||^2_{{\cal H}^\oplus_{Kin}(\lambda)}$ are measurable 
in 
particular). The operator $\MCO$ is represented on 
${\cal H}^\oplus_{Kin}(\lambda)$
as multiplication by $\lambda$, 
$\MCO (\psi(\lambda))_{\lambda\in\Rl}=
(\lambda\psi(\lambda))_{\lambda\in \Rl}$ whenever $\psi\in\mbox{dom}(\MCO)$.
The measure $\nu$ and the Hilbert spaces ${\cal H}^\oplus_{Kin}(\lambda)$ 
are not
uniquely determined but different choices give rise to unitarily equivalent
representations. Given such a choice, the scalar product on 
${\cal H}^\oplus_{Kin}(\lambda)$ is uniquely determined by that on ${\cal 
H}_{Kin}$. It follows:
\begin{Definition} \label{def3.7}
The physical Hilbert space is given by 
\be \label{3.27}
{\cal H}_{Phys}:={\cal H}^\oplus_{Kin}(0)
\ee
\end{Definition}
Notice that ${\cal H}_{Phys}$ comes automatically with a physical inner 
product, that 
is, we have simultaneously constructed the full solution space and an 
inner product on it. Moreover, one can show that if $\hat{O}_{Inv}$ is a 
strong
self-adjoint quantum Dirac observable, more precisely, if the spectral 
projections of the operators $\hat{O}_{Inv},\MCO$ commute, then 
$\hat{O}_{Inv}$ can be represented as a collection 
$(\hat{O}_{Inv}(\lambda))_{\lambda\in \Rl}$ of symmetric operators 
$\hat{O}_{Inv}(\lambda)$ on the individual ${\cal H}^\oplus_{Kin}(\lambda)$. 
If $\hat{O}_{Inv}$ is also bounded then $\hat{O}_{Inv}(\lambda)$ is even 
self-adjoint
there. Thus $\hat{O}_{Inv}(0)$ preserves ${\cal H}_{Phys}$ and the 
adjointness 
relations from ${\cal H}_{Kin}$ are transferred to ${\cal H}_{Phys}$.
Candidates for $\hat{O}_{Inv}$ are, of course, the $[\hat{O}]$ of 
(\ref{3.24}).\\
\\
Remarks:\\
1.\\
At the classical level, the immediate criticism about the {\bf Master
Constraint Proposal} is that it seems to fail to detect weak Dirac 
observables since it has vanishing Poisson brackets with every
function $O$ on the constraint surface. This criticism is wiped out by 
theorem \ref{th3.1}. 

At the quantum level there is a related criticism:
How can it be that the {\it multitude} of quantum constraints 
$\hat{C}_J$ have the same number of solutions as the the {\it single}
{\bf Quantum Master Constraint} $\MCO$ ? The answer lies in 
the functional analytic details: As long as we are just looking for 
solutions of the single equation $\MCO\psi=0$ without caring to which space 
it belongs, then it has zillions of solutions which are not solutions of 
the many equations $\hat{C}_J\psi=0$. However, the requirement that 
those solutions be normalizable with respect to the inner product 
induced, e.g. by the direct integral representation or the Rigging
Map construction, rules out those extra solutions as one can explicitly 
verify in solvable models. Alternatively, 
when constructing the physical inner product by requiring that 
a complete number of (strong)
Dirac observables (including the operators $\hat{C}_J$) be represented as 
self-adjoint operators on 
${\cal H}_{Phys}$, one finds out, in solvable models, that the solution 
space must be reduced to the simultaneous one of all constraints.
We will come back to this issue in \cite{41}.\\
2.\\
We have restricted ourselves here to first class constraints for simplicity.
But the {\bf Master Constraint Programme} can also deal with 
second 
class constraints on the classical level. On the quantum level, as 
solvable models reveal, one has to be careful with the ordering of the 
operator as otherwise there might be no solutions at all \cite{41}.\\
3.\\
We have assumed the separability of the Hilbert space because it implies 
an existence theorem about the physical Hilbert space. At an heuristic
level, without rigorous proofs, one can live with the Rigging Map 
construction, recalled in the appendix, which does not assume separability 
of the 
Hilbert space. For instance, these heuristics worked quite well for 
the spatial diffeomorphism constraint in \cite{26}.\\
4.\\
The constraint algebra of the {\bf Master Constraint Operator} is trivial 
even if the constraint algebra of the original first class constraints 
$C_J$ is not. This seems strange at first because the usual operator 
constraint quantization 
$\hat{C}_J$ needs to be supplemented by a discussion of anomaly
freeness. There is no contradiction because a non-anomalous operator 
ordering of the $\hat{C}_J$, if any, should enlarge the physical Hilbert 
space of the corresponding {\bf Master Constraint Operator}, heuristically 
given by $\MCO=\sum_J \hat{C}_J^\dagger \hat{C}_J$ where the adjoint is with 
respect to ${\cal H}_{Kin}$. Since only a non-anomalous constraint 
algebra usually results in a sufficiently large ${\cal H}_{Phys}$ 
(sufficient number of semiclassical states), the issue of anomaly freeness
has been merely translated into the size of ${\cal H}_{Phys}$. However, 
while 
physically nothing has been gained (as it should not) by the {\bf Master
Constraint Programme}, mathematically it has the huge advantage that it 
lets us carry out the quantization programme until the very end, with 
the verification of the correct semiclassical limit as the only and 
final non-trivial consistency check while simplifying the solution of the 
constraints and the construction of Dirac observables.

\section{The {\bf Master Constraint Operator} for General Relativity:\\
1. Graph-Changing Version}
\label{s4}

The purpose of this section is to sketch the quantization of the 
{\bf Master Constraint} for General Relativity. Many more details will 
follow in subsequent submissions \cite{42a}. We assume the reader to be 
familiar
with LQG at an at least introductory level and follow the notation 
of the first reference of \cite{1}.

\subsection{Classical Preliminaries}
\label{s4.1}

By $C(x)$ we mean, of course, the Lorentzian Hamiltonian constraint
of General Relativity plus all known matter in four spacetime dimensions.
For the introductory purposes of this paper we will restrict ourselves 
to the purely geometrical piece because it entails already the essential
features of the new quantization that we are about to introduce. Thus
in what follows, $C(x)$ will mean that gravitational part only.

As shown in \cite{2,3}, the smeared constraint $C(N)$ can be written 
in the following form
\ba \label{4.1}
C(N) &=& C(N)+[\beta^2+1]T(N)
\nonumber\\
C_E(N) &=& \frac{1}{\kappa}\int_\sigma N \mbox{Tr}(F\wedge e)
\nonumber\\
T(N) &=& \frac{1}{\kappa}\int_\sigma N \mbox{Tr}(K\wedge K\wedge e)
\ea
where $\kappa$ is the gravitational constant.
The $SU(2)-$valued one-forms $e$ and $K$ respectively are related to 
the intrinsic metric
$q_{ab}$ and the extrinsic curvature $K_{ab}$ of the ADM formulation 
respectively by the formulas
\be \label{4.2}
q_{ab}=\delta_{jk} e^j_a e^k_b,\;\;
K_{ab}=K_{(a}^j e^k_{b)} \delta_{jk}
\ee
so $e_a^j$ is nothing else than the cotriad.
The Ashtekar -- Barbero variables \cite{45a} of the connection 
formulation of General Relativity
are the canonically conjugate pair
\be \label{4.3}
A_a^j=\Gamma_a^j+\beta K_a^j,\;\;E^a_j=\det(e) e^a_j/\beta
\ee
where $\Gamma$ is the spin connection of $e$ and $\beta$ is called the 
Immirzi parameter \cite{45}.

In order to quantize the {\bf Master Constraint} corresponding 
to (\ref{4.1}) we proceed as in \cite{2,3} and 
use the {\bf key identities}
\ba \label{4.4}
e_a^j(x) &=& -\frac{2}{\kappa\beta} \{A_a^j(x),V(R_x)\}  
\nonumber\\
K_a^j(x) &=& -\frac{1}{\kappa\beta}\{A_a^j(x),\{C_E(1),V\}\}
\nonumber\\
V(R_x) &=& \int_\sigma \chi_{R_x}(y) d^3y \sqrt{\det(q)}(y)
\nonumber\\
C_E(1) &=& C_E(N)_{|N=1}
\ea
The quantity $V(R_x)$ is the volume of the of the open region $R_x$
which is completely arbitrary, the only condition being that $x\in R_x$.
On the other hand, $V$ is the total volume of $\sigma$ which diverges 
when $\sigma$ is not compact but this is unproblematic since $V$ appears 
only inside a Poisson bracket (its functional derivative is well-defined).
The logic 
behind (\ref{4.4}) is that $V(R)$ has a well-defined quantization 
$\hat{V}(R)$ for any region $R$
on the kinematical Hilbert space ${\cal H}_{Kin}$ of LQG \cite{46}. Thus, 
if we 
replace 
Poisson brackets by commutators divided by $i\hbar$ then we may be able
to first quantize $C_E(x)$ and then $T(x)$. 

Inserting (\ref{4.4}) into (\ref{4.1}) we obtain
\ba \label{4.5}
C_E(N) &=& -\frac{2}{\kappa^2\beta}\int_\sigma N \mbox{Tr}(F\wedge 
\{A,V\})
\nonumber\\
T(N) &=& -\frac{2}{\kappa^4\beta^3}
\int_\sigma N \mbox{Tr}
(\{A,\{C_E(1),V\}\}\wedge \{A,\{C_E(1),V\}\}\wedge \{A,V\})
\ea
Correspondingly, the {\bf Master Constraint} can be written in the form
\ba \label{4.6}
\MC &:=& \int_\sigma d^3x \frac{C(x)^2}{\sqrt{\det(q)}(x)}
=\int_\sigma d^3x (\frac{C}{\sqrt[4]{\det(q)}})(x)
\int_\sigma d^3y \delta(x,y) (\frac{C}{\sqrt[4]{\det(q)}})(y)
\\
&=& \lim_{\epsilon\to 0} (\frac{2}{\kappa\sqrt{\beta}})^4
\int_\sigma d^3x 
\mbox{Tr}([F+\frac{\beta^2+1}{(\kappa\beta)^2}\{A,\{C_E(1),V\}\}\wedge 
\{A,\{C_E(1),V\}\}]\wedge \{A,\sqrt{V_{\epsilon,.}}\})(x)
\times\nonumber\\
&& \times 
\int_\sigma d^3y \chi_\epsilon(x,y) 
\mbox{Tr}([F+\frac{\beta^2+1}{(\kappa\beta)^2}\{A,\{C_E(1),V\}\}\wedge 
\{A,\{C_E(1),V\}\}]\wedge \{A,\sqrt{V_{\epsilon,.}}\})(y)
\nonumber
\ea
Here $\chi_\epsilon(x,y)$ is any one-parameter family of, not 
necessarily smooth, functions
such that $\lim_{\epsilon\to 0} \chi_\epsilon(x,y)/\epsilon^3=\delta(x,y)$
and $\chi_\epsilon(x,x)=1$. Moreover,
\be \label{4.7}
V_{\epsilon,x}:=\int_\sigma d^3y \chi_\epsilon(x,y) \sqrt{\det(q)}(y)
\ee
Thus, in the last line of (\ref{4.6}) we have performed a convenient 
point split. 

Readers familiar with \cite{2,3,4,5,6,7,8,9} already recognize
that the integrands of the two integrals in (\ref{4.6}) are {\it 
precisely} those of \cite{3}, the only difference being that the last
factor in the wedge product is given by $\{A,\sqrt{V_{\epsilon,.}}\}$
rather than $\{A,V\}$ which comes from the additional factor of 
$(\root 4 \of{\det(q)})^{-1}$ in our point-split expression. Thus we 
proceed exactly as in \cite{3} and introduce a partition $\cal P$
of $\sigma$ into cells $\Box$, splitting both integrals into sums
$\int_\sigma=\sum_{\Box\in {\cal P}}$. Assigning to each cell $\Box$
an interior point $v(\Box)$, in the infinite refinement limit 
${\cal P}\to\sigma$ in which all the cells collapse to a single point and 
those points fill all of $\sigma$ we may replace (\ref{4.6}) by the
limit of the following double Riemann sum
\ba \label{4.8}
\MC  
&=& \lim_{\epsilon\to 0} \lim_{{\cal P}\to\sigma} 
\sum_{\Box,\Box'\in {\cal P}} \chi_\epsilon(v(\Box),v(\Box'))
(\frac{2}{\kappa\sqrt{\beta}})^4 
\times \nonumber\\
&&\times \int_\Box d^3x 
\mbox{Tr}([F+\frac{\beta^2+1}{(\kappa\beta)^2}\{A,\{C_E(1),V\}\}\wedge 
\{A,\{C_E(1),V\}\}]\wedge \{A,\sqrt{V_{\epsilon,.}}\})(x)
\times\nonumber\\
&& \times 
\int_{\Box'} d^3y 
\mbox{Tr}([F+\frac{\beta^2+1}{(\kappa\beta)^2}\{A,\{C_E(1),V\}\}\wedge 
\{A,\{C_E(1),V\}\}]\wedge \{A,\sqrt{V_{\epsilon,.}}\})(y)
\ea
Notice that the limit of the Riemann sum is independent of the way 
we refine the partition.

Now the integrals over $\Box,\Box'$ in (\ref{4.8})
are of the type of functions
that can be quantized by the methods of \cite{2,3,4,5,6,7,8,9}.
Before we do that we recall why the limit ${\cal P}\to\sigma$ 
does not exist in the sense of operators on the kinematical Hilbert space 
and forces us to define the operator (or rather a quadratic form) directly
on the diffeomorphism invariant Hilbert space ${\cal H}_{Diff}$ \cite{26}.
This makes sense for $\MC$ in contrast to $C(N)$ since $\MC$ is a 
diffeomorphism invariant function.

\subsection{Spatially Diffeomorphism Invariant Operators}
\label{s4.3}

\begin{Lemma} \label{la4.1} ~~~\\
Let $Q$ be a spatially diffeomorphism invariant 
quadratic form on 
${\cal H}_{Kin}$ whose form domain contains the smooth cylindrical 
functions 
Cyl$^\infty$. Let $Q_{s,s'}:=Q(T_s,T_{s'})$ where $s$ denotes a spin
network (SNW) label and $T_s$ the corresponding spin network function 
(SNWF). Then a 
necessary condition for $Q$ to be the quadratic form of a spatially 
diffeomorphism invariant operator densely defined on Cyl$^\infty$ is that 
$Q_{s,s'}=0$ whenever $\gamma(s)\not=\gamma(s')$ where $\gamma(s)$ denotes 
the graph label of $s$.
\end{Lemma}
Proof of lemma \ref{la4.1}:\\
Recall the unitary representation 
$\hat{U}:\;\mbox{Diff}^\omega(\sigma)\to {\cal B}({\cal H}_{Kin})$ of the 
group of analytic diffeomorphisms of $\sigma$ on the kinematical Hilbert
space defined by 
$\hat{U}(\varphi)p_\gamma^\ast f_\gamma=
p_{\varphi(\gamma)}^\ast f_\gamma$ where $p_\gamma$ is the restriction 
of a (generalized) connection $A\in\ab$ to the edges $e\in E(\gamma)$
and $f_\gamma;\;SU(2)^{|E(\gamma)|}\to \Cl$. A 
spatially diffeomorphism invariant quadratic form then is defined by 
$Q(U(\varphi)f,U(\varphi)f')=Q(f,f')$ for all $f,f'\in \mbox{dom}(Q)$.

Consider any $\gamma(s)\not=\gamma(s')$ where $\gamma(s)$ is the graph 
underlying the SNW $s$. Then we find a countably 
infinite number of analytic diffeomorphisms $\varphi_n,\;n=0,1,2,..$ such 
that $T_{s'}$ is invariant but $T_{s_n}=U(\varphi_n) T_s$ are mutually
orthogonal spin network states (we have set $\varphi_0=\mbox{id}$;
interchange $s,s'$ if $T_s=1$). 
Suppose now that $Q$ is the quadratic form
of a spatially diffeomorphism invariant operator 
$\hat{O}$ on ${\cal H}_{Kin}$, that is, 
$\hat{U}(\varphi)\hat{O}\hat{U}(\varphi)^{-1}=\hat{O}$ for all 
$\varphi\in \mbox{Diff}^\omega(\sigma)$, $Q(f,f')=<f,\hat{O} f'>$, densely 
defined on Cyl$^\infty$.
Then 
\be \label{4.11}
||\hat{O}T_s'||^2=\sum_s |Q_{s,s'}|^2\ge
\sum_{n=0}^\infty |Q_{s_n,s'}|^2=|Q_{s,s'}|^2[\sum_{n=0} 1]
\ee
diverges unless $Q_{s,s'}=0$.\\
$\Box$\\
We conclude that diffeomorphism invariant operators which 
are graph 
changing cannot exist on ${\cal H}_{Kin}$. The only diffeomorphism 
invariant 
operators which can be defined on ${\cal H}_{Kin}$ must not involve the 
connection $A$, they are defined purely in terms of $E$. The total volume 
of $\sigma$ is an example for such an operator. 
Since $\MC$ involves 
the curvature $F$ of $A$, the operator $\MCO$ cannot be defined 
on ${\cal H}_{Kin}$ unless one uses an ad hoc procedure as in section 
\ref{s5}. The way out, as noticed in \cite{2,3,4}, is to define 
$\MCO$ not on ${\cal H}_{Kin}$ but on ${\cal H}_{Diff}$. The effect of 
this is, roughly speaking, that all the terms in the infinite sum in 
(\ref{4.11}) are eqivalent under diffeomorphisms, hence we also need only 
one of them, whence the infinite sum becomes finite.

\subsection{Diffeomorphism Invariant Hilbert Space}
\label{s4.4}

Recall the definition 
of ${\cal H}_{Diff}$. A diffeomorphism invariant distribution $l$ is a,
not necessarily continuous, linear functional on 
$\Phi_{Kin}:=$Cyl$^\infty$ (that is, an element of $\Phi_{Kin}^\ast$) such 
that
\be \label{4.12}
l(\hat{U}(\varphi)f)=l(f) \;\;\forall \;\; f\in \mbox{Cyl}^\infty,\;
\varphi\in\mbox{Diff}^\omega(\sigma)
\ee
Since the finite linear span of SNWF's is dense in Cyl$^\infty$
it suffices to define $l$ on the SNWF, hence $l$ can formally be written 
in the form
\be \label{4.13}
l(.)=\sum_s l_s <T_s,.>_{Kin}
\ee
for complex valued coefficients which satisfy $l_s=l_{s'}$ whenever
$T_{s'}=\hat{U}(\varphi) T_s$ for some diffeomorphism $\varphi$. 
In other words, the coefficients only depend on the orbits 
\be \label{4.14}
[s]=\{s';\; T_{s'}=\hat{U}(\varphi) T_s \mbox{ for some } \varphi
\in \mbox{Diff}^\omega(\sigma)\}
\ee
Hence the distributions 
\be \label{4.15}
b_{[s]}=\sum_{s'\in [s]} <T_{s'},.>_{Kin}
\ee
play a distinguished role since 
\be \label{4.16}
l(.)=\sum_{[s]} l_{[s]} b_{[s]}
\ee
The Hilbert space ${\cal H}_{Diff}$ is now defined as the completion 
of the finite linear span of the $b_{[s]}$ in the scalar product 
obtained by declaring 
the $b_{[s]}$ to be an orthonormal basis, see \cite{26} for a detailed 
derivation through the procedure of refined algebraic quantization
(group avaraging) \cite{44}. More 
explicitly, the scalar product $<.,.>_{Diff}$ is defined on the basis 
elements by
\be \label{4.17}
<b_{[s]},b_{[s']}>_{Diff}=
\overline{b_{[s]}(T_{s'})}=
b_{[s']}(T_{s})=\chi_{[s]}(s')=\delta_{[s],[s']}
\ee
and extended by sesquilinearity to the finite linear span of the 
$b_{[s]}$.

In this paper we will actually modify the Hilbert space ${\cal H}_{Diff}$
as follows: One might think that ${\cal H}_{Diff}$ as defined is 
separable, however, this is not the case: As shown explicitly in 
\cite{38}, since the diffeomorphism group at a given point reduces 
to $GL_+(3,\Rl)$, for vertices of valence five or higher there are 
continuous, diffeomorphism invariant parameters associated with such 
vertices. More precisely, we have the following: Given an $n-$valent 
vertex $v$ of a graph $\gamma$, consider all its  ${n \choose 3}$  
triples $(e_1,e_2,e_3)$ of edges incident at it. With each triple we 
associate a degeneracy type $\tau(e_1,e_2,e_3)$ taking six values 
depending on whether the tangents of the respective edges at $v$ are 
0) linearly independent, 
1) co-planar but no two tangents are co-linear,
2a) co-planar and precisely one pair of tangents is co-linear but
the corresponding edges are not analytic continuations of each other, 
2b) co-planar and precisely one pair of tangents is co-linear where 
the corresponding edges are analytic continuations of each other, 
3a) co-linear but for no pair the corresponding edges are analytic 
continuations of each other,  
3b) co-linear and precisely for one pair the corresponding edges are 
analytic continuations of each other.
Notice that an analytic diffeomorphism or a reparameterization cannot 
change the degenaracy type of any triple. We could refine the 
classification of degeneracy types by considering also the derivatives of 
the edges of order $1<k<\infty$ (we just considered $k=1,\infty$) in which 
case we would associate more discrete diffeomorphism invariant 
information with a vertex of valence $n$ but for our purposes this will 
sufficent. By the degeneracy type of a graph we will mean the collection 
of degeneracy types of each of the triples of all vertices.

Given a SNW $s$ with 
$\gamma(s)=\gamma$ we now mean by $\{s\}$ the set of all $s'$ such 
that $[s]$ and $[s']$ differ at most by a different value of continuous
moduli but not by a degeneracy type for any triple of edges for any 
vertex. In other words, ${s}={s'}$ whenever $\gamma(s),\gamma(s')$ are 
{\it ambient isotopic} up to the degeneracy type. That is, $\gamma(s)$
can be deformed into $\gamma(s')$ by a smooth one-parameter family of 
analytic maps $f_t:\;\sigma\to\sigma,\; t\in [0,1]$ with analytic inverse 
such that 
$f_0(\gamma(s))=\gamma(s),\;f_1(\gamma(s))=\gamma(s')$ without changing 
the degeneracy type of $\gamma(s)$. If $\gamma(s)$ has at most four-valent 
vertices then $\{s\}=[s]$ but otherwise these two classes are different.
The class $\{s\}$ depends only on discrete labels and if we define 
\be \label{4.18}
b_{\{s\}}(.)=\sum_{s'\in \{s\}} <T_{s'},.>_{Kin}
\ee
and 
\be \label{4.19}
<b_{\{s\}},b_{\{s'\}}>_{Diff}:=
\overline{b_{\{s\}}(T_{s'})}=
b_{\{s'\}}(T_{s})=\chi_{\{s\}}(s')=\delta_{\{s\},\{s'\}}
\ee
then ${\cal H}_{Diff}$ becomes separable. The passage from (\ref{4.16}),
(\ref{4.17}) (diffeomorphism invariant states) to (\ref{4.18}), 
(\ref{4.19}) (ambient isotopy modulo degeneracy type invariant states)  
may seem as a drastic step because ambient isotopy is not induced by the 
symmetry group. On the other hand, for graphs with at most four valent 
vertices, which are the most important ones for semi-classical 
considerations since the dual graph of any simplicial cellular 
decomposition of a 
manifold is four-valent, there is no difference. Moreover, any 
$b_{\{s\}}$ is a (possibly uncountably infinite) linear combination 
over of the $b_{[s]}$ by summing over the corresponding continuous 
moduli. There might be other ways to get rid of the non-separabilty,
for example by introducing a measure on the {\it Teichm\"uller}-like space 
of continuous moduli, but this goes beyond the scope of the present 
paper.

\subsection{Regularization and Quantization of the {\bf Master Constraint}}
\label{s4.6}

We now have recalled all the tools to quantize $\MC$ itself. As we have 
explained in detail, this will be possible only on ${\cal H}_{Diff}$.
Moreover, we must
construct an operator $\MCO$ (or rather a dual operator $\MCO'$ on 
${\cal H}_{Diff}$) which is positive in order that it really enforces the 
constraint $C(x)=0$ for all $x\in\sigma$. The following strategy 
allows us to guarantee this by construction:\\
\\
At given $\epsilon,{\cal P}$ the basic building blocks of (\ref{4.8})
are the integrals
\be \label{4.20}
C_{\epsilon,{\cal P}}(\Box)
=\int_\Box d^3x 
\mbox{Tr}([F+\frac{\beta^2+1}{(\kappa\beta)^2}\{A,\{C_E(1),V\}\}\wedge 
\{A,\{C_E(1),V\}\}]\wedge \{A,\sqrt{V_{\epsilon,.}}\})(x)
\ee
In \cite{3} the following strategy for the quantization of operators of the
type of (\ref{4.20}) was developed:
\begin{itemize}
\item[1.] {\it Regulated Operator on ${\cal H}_{Kin}$}\\
Replacing Poisson brackets by commutators divided by $i\hbar$ and 
functions by their corresponding operators, we obtain a regulated operator
$\hat{C}^\dagger_{\epsilon,{\cal P}}(\Box)$ which, together with its 
adjoint, is densely defined on the invariant domain 
$\Phi_{Kin}:=$Cyl$^\infty$.
\item[2.] {\it Dual Reguated Operator on $\Phi_{Kin}^\ast$}\\
Let $\Phi^\ast_{Kin}$ be the algebraic dual of $\Phi_{Kin}$, then we may 
define a dual
regulated operator $\hat{C}'_{\epsilon,{\cal P}}(\Box)$ on 
$\Phi^\ast_{Kin}$ by 
the formula
\be \label{4.21}
(\hat{C}'_{\epsilon,{\cal P}}(\Box) l)[f]
:=l[\hat{C}^\dagger_{\epsilon,{\cal P}}(\Box) f]
\ee
\item[3.] {\it Taking $\epsilon\to 0$}\\
One regularization step is to interchange the limits in (\ref{4.8}),
hence we take $\epsilon\to 0$ at finite $\cal P$. This enforces 
that the douple sum collapses to a single one since 
$\lim_{\epsilon\to 0}\chi_\epsilon(v(\Box),v(\Box'))=
\delta_{\Box,\Box'}$, moreover, due to the particulars of the volume operator
also the limit $\epsilon\to 0$ of $\hat{C}'_{\epsilon,{\cal P}}(\Box)$
exists in the topology of pointwise convergence on $\Phi_{Kin}^\ast$, 
resulting in $\hat{C}'_{{\cal P}}(\Box)$.
\item[4.] {\it Refinement Limit}\\
The limit ${\cal P}\to \sigma$ does not exist in the 
topology of
pointwise convergence on all of $\Phi_{Kin}^\ast$ but on the subspace 
$(\Phi_{Kin}^\ast)_{Diff}$ of diffeomorphism invariant distributions. The 
limit 
defines an element of $\Phi_{Kin}^\ast$ again but it is not an element of 
$(\Phi_{Kin}^\ast)_{Diff}$, hence it does not leave that space invariant.
\end{itemize}
In \cite{3} these steps were used in order to define a smeared Hamiltonian
constraint operator densely defined on $\Phi_{Kin}$ by introducing a new 
kind of
operator topology which involves $(\Phi_{Kin}^\ast)_{Diff}$. The 
definition 
of the resulting operator involves the axiom of choice, hence it exists but 
cannot be written down explicitly, only its dual action on 
$(\Phi_{Kin}^\ast)_{Diff}$ can be written explicitly. It is at this 
point that we 
use a different strategy. Namely we propose the following heuristic 
expression for a quadratic form 
\be \label{4.22}
Q_{\MC}(l,l'):=
\lim_{{\cal P}\to\sigma} \sum_{\Box \in {\cal P}} 
(\frac{2}{\kappa\sqrt{\beta}})^4 
<\hat{C}'_{{\cal P}}(\Box) l,\hat{C}'_{{\cal P}}(\Box) l'>_{Diff}
\ee
on the dense subspace of ${\cal H}_{Diff}$ defined by the finite linear 
span of the $b_{\{s\}}$. 

The immediate problem with (\ref{4.22}) is that the objects 
$\hat{C}'_{{\cal P}}(\Box) l$ are elements of $\Phi_{Kin}^\ast$ but in 
general 
not of $(\Phi_{Kin}^\ast)_{Diff}$, hence the scalar product on ${\cal 
H}_{Diff}$
in the last line of (\ref{4.22}) is ill-defined. 
However, as we will show in detail in \cite{42a}, the scalar product {\it 
does} become well-defined in the limit ${\cal P}\to\sigma$ for the same 
reason that the classical expression (\ref{4.8}) becomes diffeomorphism 
invariant only in the limit ${\cal P}\to\sigma$. The precise proof
of this fact is lengthy and goes beyond the scope of the present paper,
hence we will restrict ourselves here to a heuristic derivation 
which provides a shortcut to the final formula. (Alternatively, one may 
view the procedure below as {\it part of the of the regularization}).

The shortcut is based on a {\bf resolution of identity trick}:\\
Making use of the fact that the $b_{\{s\}}$
form an orthonormal basis of ${\cal H}_{Diff}$ and pretending that
the $\hat{C}'_{{\cal P}}(\Box) l$ are elements of ${\cal H}_{Diff}$ 
we can insert an identity 
\be \label{4.23}
Q_{\MC}(l,l'):=
\lim_{{\cal P}\to\sigma} \sum_{\Box \in {\cal P}} 
(\frac{2}{\kappa\sqrt{\beta}})^4 
\sum_{\{s\}}\;\;
<\hat{C}'_{{\cal P}}(\Box) l,b_{\{s\}}>_{Diff} \;\;
<b_{\{s\}},\hat{C}'_{{\cal P}}(\Box) l'>_{Diff} 
\ee
Still pretending that the $\hat{C}'_{{\cal P}}(\Box) l$ are elements of 
${\cal H}_{Diff}$ we may now use the definition of the scalar product
on ${\cal H}_{Diff}$ and arrive at the, now meaningful, 
expression
\be \label{4.24}
Q_{\MC}(l,l'):=\sum_{\{s\}}
\lim_{{\cal P}\to\sigma} \sum_{\Box \in {\cal P}} 
(\frac{2}{\kappa\sqrt{\beta}})^4 \;\;
(\hat{C}'_{{\cal P}}(\Box) l)[T_{s_0(\{s\})}] \;\;
\overline{(\hat{C}'_{{\cal P}}(\Box) l')[T_{s_0(\{s\})}]}
\ee
where we have chosen suitable representatives $s_0(\{s\})\in \{s\}$
and have interchanged the sum $\sum_{\{s\}}$ with the limit
${\cal P}\to \sigma$ which is again part of the regularization.
Now one of the properties of the operators $\hat{C}'_{{\cal P}}(\Box)$ 
that were proved in \cite{3} is that the dependence of 
$(\hat{C}'_{{\cal P}}(\Box) l)[T_{s_0(\{s\})}]$ on the representative 
$s_0(\{s\})$ reduces to the question of how many vertices of 
$\gamma(s_0(\{s\}))$ of which vertex type are contained in $\Box$. For each 
vertex contained 
in $\Box$ we obtain a contribution which no longer depends on the 
representative but only on the diffeomorphism class of the vertex
(or the ambient isotopy class up to the degenracy type), that is, its 
vertex type. Let us write this as 
\be \label{4.25}
(\hat{C}'_{{\cal P}}(\Box) l)[T_{s_0(\{s\})}]
=\sum_{v\in V(\gamma(s_0(\{s\})))\cap \Box}
(\hat{C}'_{{\cal P}}(v) l)[T_{s_0(\{s\})}]
\ee
where, thanks to the diffeomorphism (or ambient isotopy up to 
degeneracy) invariance of $l$ the numbers $(\hat{C}'_{{\cal P}}(v) 
l)[T_{s_0(\{s\})}]$
depend only on the vertex type of $v$ and are, in this sense,
diffeomorphism (or ambient isotopy up to degeneracy type) invariant.

Now, in the limit ${\cal P}\to\infty$ it is clear that no matter 
how that limit is reached and no matter which representative was chosen,
each $\Box$ contains at most one vertex of $\gamma(s_0(\{s\}))$. 
Moreover, recall that by definition in \cite{3} the partition is refined 
depending on $s_0(\{s\})$ in such a way that its topology is constant in 
the neighbourhood of any vertex of $\gamma(s_0(\{s\}))$ for 
sufficiently fine partition. This state dependent
regularization is justified by the fact that the classical Riemann 
sum converges to the same integral no matter how the partition is refined.
More precisely, the state dependent refinement limit
is such that eventually for each vertex $v\in V(\gamma(s_0(\{s\}))$
there is precisely one cell $\Box_v$ which contains $v$ as an interior
point and the partition is refined in such a way that $\Box_v\to \{v\}$
while it is arbitrary for any cell which does not contain a vertex. 

It follows that for each class $\{s\}$ and any representative 
$s_0(\{s\})$ the partition $\cal P$ will eventually be so fine that
the numbers 
\be \label{4.26}
(\hat{C}'_{{\cal P}}(\Box_v) l)[T_{s_0(\{s\})}]
=:(\hat{C}'_{{\cal P}}(v) l)[T_{s_0(\{s\})}]
\ee
do not change any more as $\Box_v\to \{v\}$. Hence (\ref{4.24})
can eventually be written in the form 
\ba \label{4.27}
Q_{\MC}(l,l') \
&=& \sum_{\{s\}} \lim_{{\cal P}\to\sigma} \sum_{v\in V(\gamma(s_0(\{s\})))} 
(\frac{2}{\kappa\sqrt{\beta}})^4 \;\;
(\hat{C}'_{{\cal P}}(\Box_v) l)[T_{s_0(\{s\})}] \;\;
\overline{(\hat{C}'_{{\cal P}}(\Box_v) l')[T_{s_0(\{s\})}]}
\nonumber\\
&=& \sum_{\{s\}} \lim_{{\cal P}\to\sigma} \sum_{v\in V(\gamma(s_0(\{s\})))} 
(\frac{2}{\kappa\sqrt{\beta}})^4 \;\;
(\hat{C}'_{{\cal P}}(v) l)[T_{s_0(\{s\})}]\;\;
\overline{(\hat{C}'_{{\cal P}}(v) l')[T_{s_0(\{s\})}]}
\nonumber\\
&=& \sum_{\{s\}} \sum_{v\in V(\gamma(s_0(\{s\})))} 
(\frac{2}{\kappa\sqrt{\beta}})^4 \;\;
(\hat{C}'_{{\cal P}}(v) l)[T_{s_0(\{s\})}]\;\;
\overline{(\hat{C}'_{{\cal P}}(v) l')[T_{s_0(\{s\})}]}
\ea
where in the last line we could finally take the limit ${\cal P}\to \sigma$.
The beauty of (\ref{4.24}) is that it really is independent of the 
choice of the representative $s_0(\{s\})$ because for any representative 
we get the same number of vertices of each vertex type, hence the sum of 
the contributions does not change under change of representative. Thus,
the {\bf axiom of choice is no longer necessary to compute (\ref{4.27}).}

The quadratic form (\ref{4.27}) is positive by 
inspection and since, in the sense we just described, it comes from 
an operator, chances a good that it can be closed and hence
determines a unique self-adjoint operator $\MCO$ on ${\cal H}_{Diff}$.
We will examine this in a future publication \cite{42a}.
Notice that it is neither necessary nor possible to check whether 
$Q_{\MC}$ defines a spatially diffeomorphism invariant quadratic form 
since it is already defined on ${\cal H}_{Diff}$. We could check this 
only if we would have a quadratic form on the kinematical Hilbert space
${\cal H}_{Kin}$ but, as we explained, this is impossible for a graph 
changing 
operator. The only remnant of such a check is the independence of 
(\ref{4.27}) of the choice of representatives $s_0(\{s\})$ which we just did.

The expression (\ref{4.27}), seems to be hard to compute due to the 
infinite sum $\sum_{\{s\}}$, it even looks divergent. However, this is 
not the case: Let $\Phi_{Diff}$ be the dense (in ${\cal H}_{Diff}$) subset 
of $(\Phi_{Kin}^\ast)_{Diff}$
consisting of the finite linear combinations of the $b_{\{s\}}$.
Then for given $l,l'\in \Phi_{Diff}$ there 
are always only a finite number of terms that contribute to (\ref{4.27}).
This also justifies the interchange of the limit ${\cal P}\to \sigma$ and 
the sum $\sum_{\{s\}}$ performed in (\ref{4.27})
Hence the dense subspace $\Phi_{Diff}\subset {\cal H}_{Diff}$ certainly 
belongs to the {\it form domain} of the positive quadratic form $Q_{\MC}$.
Notice that this reasoning would hold also if we would only stick with 
the non-separable ${\cal H}_{Diff}$, separability is required only 
for the direct integral decomposition of ${\cal H}_{Diff}$.\\
\\
Remarks:\\
1.\\
A serious criticism spelled out in \cite{15} is that the action of the 
Hamiltonian constraint of \cite{2,3,4} is ``too local" in the sense that
it does not act at the vertices that it creates itself, so in some sense 
information does not propagate. While this is inconclusive because 
the constraint certainly acts everywhere, it is still something to
worry about. The reason for why the action at those vertices had to be 
trivial was that otherwise the Hamiltonian constraint would be anomalous.
Now the {\bf Master Constraint} is not subject to any non-trivial algebra
relations and hence {\bf increases our flexibility in the way it acts
at vertices that it creates itself}, thus at least relaxing the worries
spleeld out in \cite{15}. In \cite{42a} we will come back to this issue.\\  
2.\\
We have stressed that we would like to prove that $Q_{\MC}$ has a closure
in order to rigorously grant existence of $\MCO$. 
At a heuristic level (group averaging, Rigging Map) one can even live just 
with the quadratic form: Consider the dense subspace 
$\Phi_{Diff}$ of ${\cal H}_{Diff}$ and its algebraic dual 
$\Phi_{Diff}^\ast$ (not to be 
confused with the subset $(\Phi_{Kin}^\ast)_{Diff}$ of kinematical 
algebraic 
distributions $\Phi_{Kin}^\ast$). Then typical elements of the physical
subspace 
$(\Phi_{Diff}^\ast)_{Phys}$ of $\Phi^\ast_{Diff}$ will 
be of the form $l_{Phys}=\sum_{\{s\}} z_{\{s\}} <b_{\{s\}},.>_{Diff}$
with complex valued coefficients $z_{\{s\}}$ and one would impose the 
condition 
\be \label{4.28}
(\MCO' l_{Phys})[F]:=
l_{Phys}[\MCO F]:=\sum_{\{s\}} z_{\{s\}} 
Q_{\MC}(b_{\{s\}},F)=0
\ee
for all $F\in \Phi_{Diff}$. Notice that this condition constructs possibly a 
solution space but not an inner product for which we would need at 
least a Rigging Map. If there is a subset 
$\Phi_{Phys}\subset (\Phi_{Diff}^\ast)_{Phys}$ of would-be 
analytic vectors for $\MCO$ then one could try to define such a map 
as
\ba \label{4.29}
&& <\eta_{Phys}(F),\eta_{Phys}(F')>_{Phys}
:= \int_\Rl \frac{dt}{2\pi} <F',e^{it\MCO} F>_{Diff}
\nonumber\\
&:=& \int_\Rl \frac{dt}{2\pi} 
\sum_{n=0}^\infty \frac{(it)^n}{n!}<F', \MCO^n F>_{Diff}
\nonumber\\
&:=& \int_\Rl \frac{dt}{2\pi} 
\sum_{n=0}^\infty \frac{(it)^n}{n!}\sum_{\{s_1\},..,\{s_n\}}
Q_{\MC}(F',b_{\{s_1\}})
Q_{\MC}(b_{\{s_1\}},b_{\{s_2\}})..Q_{\MC}(b_{\{s_n\}},F)
\ea
But certainly the existence of these objects is far from granted.

\section{The {\bf Master Constraint Operator} for General Relativity:\\
2. Non-Graph-Changing Version}
\label{s5}

As we have explained in detail in section \ref{s4.3}, a {\bf Master 
Constraint Operator} which changes the graph underlying a spin-network
state is incompatible with a diffeomorphism invariant operator. Now the 
original motivation \cite{2} for having a graph-changing Hamiltonian 
constraint was to have an anomaly free constraint algebra among the 
smeared Hamiltonian constraints $\hat{C}(N)$. This motivation is {\bf 
void} with respect to $\MCO$ since there is only one $\MCO$ so there 
{\bf cannot be any anomaly} (at most in the sense that ${\cal H}_{Phys}$
is too small, that is, has an unsufficient number of semiclassical 
states). It is therefore worthwhile thinking about 
about a version of the $\MCO$ which does not change the graph {\bf and 
therefore can be defined on the kinematical Hilbert space}. That 
this is indeed possible, even for diffeomorphism invariant operators 
which are positive 
was demonstrated in \cite{50}: Essentially one must define $\MCO$
in the spin-network basis and for each $T_s$ and each $v\in \gamma(s)$ 
one must invent a unique diffeomorphism covariant prescription for how 
to choose loops 
as {\bf parts of the already existing graph $\gamma(s)$}. In \cite{50}
we chose the {\it minimal loop prescription}: 
\begin{Definition} \label{def5.1} ~~~~\\
Given a graph $\gamma$
and a vertex $v\in V(\gamma)$ and two different edges $e,e'\in E(\gamma)$ 
starting in $v$, a loop $\alpha_{\gamma,v,e,e'}$ within $\gamma$ starting 
along $e$ and 
ending along $(e')^{-1}$ is said to be minimal provided that there 
is no other loop with the same properties and fewer edges of $\gamma$ 
traversed. 
\end{Definition}
Notice that
the notion of a minimal loop does not refer to a background metric and 
is obviously covariant. Actually it is a notion of {\bf algebraic graph 
theory} \cite{51} since it does not refer to any knotting (embedding). 
Thus, it might be possible to make the quantum dynamics look more 
combinatorical.

Given the data $\gamma,v,e,e'$, there may be more than one minimal loop
but there is at least one (we consider only closed graphs due to gauge 
invariance). We denote by $L_{\gamma,v,e,e'}$ the set of minimal 
loops associated with the data $\gamma,v,e,e'$ (It might be possible 
to make $\alpha_{\gamma,v,e,e'}$ unique by asking more properties but 
we could not think of a prescription which is also covariant). If
$L_{\gamma,v,e,e'}$ has more than one element then the corresponding
{\bf Master Constraint Operator} averages over the finite number of
elements of $L_{\gamma,v,e,e'}$, see \cite{50} for details. 

The advantage of having a non-graph changing {\bf Master Constraint 
Operator} is that one can quantize it directly as a positive 
operator on ${\cal H}_{Kin}$ and check its semi-classical properties by 
testing it with the semi-classical tools developed in 
\cite{29,30,31,32,33,34}, to the best of our knowledge currently the only 
semi-classical states, in a representation supporting the holonomy -- 
flux algebra, 
for non-Abelean gauge theories for which semi-classical properties were 
proved. Notice that the original problems of these states mentioned 
in \cite{29}, namely that they have insufficient semiclassical
properties as far as holonomy and area operators are concerned, were 
resolved in \cite{34}: Indeed, they are designed to display semi-classical
properties with respect to operators which come from functions on phase
space involving {\bf three-dimensional integrals} over $\sigma$ rather than 
one -- or 
two-dimensional integrals, provided they are not graph-changing. Hence 
they are suitable for our diffeomorphism invariant operators, in 
particular $\MCO$, when quantized without changing the graph. That
then correct semiclassical expectation values and small fluctuations 
are indeed obtained has been verified explicitly for diffeomorphism 
invariant operators for matter QFT's coupled to LQG in the second 
reference of \cite{50}. Hence we are optimistic that this is possible 
here as well, especially due to some recent progress concerning the 
spectral analysis of the volume operator \cite{52}. 
Details will be published in \cite{53}. 

The disadvantage of a non-graph-changing operator is that it uses 
a prescription like the above minimal loop prescription as an ad hoc 
quantization step. While it is motivated by the more fundamental 
quantization procedure of the previous section and is actually not 
too drastic a modification thereof for sufficiently fine graphs, the 
procedure of the previous section should be considered as more 
fundamental. 
Maybe one could call the operator as formulated in this 
section an {\bf effective operator} since it presumably reproduces all 
semiclassical properties. On the other hand, since it defines a positive 
self-adjoint operator by construction which is diffeomorphism invariant
and presumably has the correct classical limit we can exploit the full
power of techniques that comes with ${\cal H}_{Diff}$ in order to solve 
the constraint. Even better than that, since the operators 
$\MCO$ and $\hat{U}(\varphi)$ commute for all $\varphi\in 
\mbox{Diff}^\omega(\sigma)$ we can even solve the constraint $\MCO$ 
at the kinematical level in the subspace ${\cal H}_{Kin,\gamma}\subset 
{\cal H}_{Kin}$ spanned by spin network states over the graph $\gamma$,
for each $\gamma$ separately. This space is 
definitely separable and hence we do
not need to invoke the additional averaging to produce the states 
$b_{\{s\}}$ when we solve the spatial diffeomorphism constraint in a 
second step. (Notice, however, that while $\MCO$ and $\hat{U}(\varphi)$
commute, one cannot simply map the solutions to $\MCO=0$ by the 
map $\eta(T_s)=b_{\{s\}}$ because $\eta$ is not really a Rigging map,
it is a Rigging map on each sector ${\cal H}_{Kin,\{s\}}$, the closure of 
the finite linear span of the $T_{s'},\;s'\in \{s\}$ but the averaging 
weights for each sector are different due to graph symmetries which 
depend on the spin labels, see 
\cite{26} for details). Thus, although we work in the continuum, 
for each graph $\gamma$ we have to solve essentially a problem in 
Hamiltonian lattice gauge theory when solving the {\bf Master Constraint}
before the spatial diffeomorphism constraint.

\section{Further Directions and Connection with Spin Foam Models}
\label{s6}

We finish the paper with remarks and further ideas about
applications of the {\bf Master Constraint Procedure}:
\begin{itemize}
\item[i)] {\it {\bf Extended Master Constraint}}\\
Up to now we have treated the spatial 
diffeomorphism constraint and the Hamiltonian constraint on rather unequal
footing: The diffeomorphism constraint was solved in the usual way by 
imposing $C_a(x)=0\;\forall \;x\in\sigma$ while the Hamiltonian constraint
was solved by the {\bf Master Constraint Method}. This is of course 
natural because we already have a close to complete framework for the 
diffeomorphism constraint \cite{26} so we may leave things as they are 
with respect to the spatial diffeomorphism constraint. One the other hand
it is questional why smooth, even analytic diffeomorphisms should play a 
fundamental role in LQG which seems to predict a discrete structure 
at Planck scale as the spectrum of the length, area and volume operators 
reveal. Hence, one might want to consider a different approach to the 
solution of the spatial diffeomorphism constraint. 

The {\bf Master
Constraint Programme} allows us to precisely do that and to treat 
all constraints on equal footing. Consider the {\bf Extended Master
Constraint}
\be \label{6.1}
\MC_E:=\frac{1}{2}\int_\sigma d^3x \frac{C(x)^2+q^{ab}(x) C_a(x) 
C_b(x)}{\sqrt{det(q)(x)}}
\ee
Obviously $\MC_E=0$ if and only if $C(x)=C_a(x)=0$ for all $a=1,2,3;\;
x\in \sigma$, hence the general theory of section \ref{s3} applies.
Notice that while $C_a(x)$ cannot be quantized as the self-adjoint 
generator of one parameter unitary subgroups of the representation 
$\hat{U}(\varphi)$ of the spatial diffeomorphism group on ${\cal H}_{Kin}$
(since the representation is not strongly continuous), the general 
theorems of \cite{7}
show that (\ref{6.1}) has a chance to be quantized as a positive 
self-adjoint operator on ${\cal H}_{Diff}$ by the methods of section
\ref{s4} or on ${\cal H}_{Kin}$ by the methods of section \ref{s5}. Since 
(\ref{6.1}) is to {\it define} ${\cal H}_{Phys}$ is a single stroke and 
to define spatial diffeomorphism invariance in a new way, we do not really
have Diff$^\omega(\sigma)$ any more and thus the method of section
\ref{s5} is preferred. Moreover, the obstruction mentioned in section
\ref{s4.3} to having a graph changing diffeomorphism invariant operator 
on ${\cal H}_{Kin}$ is not present any more, again because 
Diff$^\omega(\sigma)$
no longer exists at the quantum level although it should be recovered 
on large scales. Thus $\MCO_E$ may now be defined as a graph changing 
operator ${\cal H}_{Kin}$. Finally, we even have the flexibility to
change ${\cal H}_{Kin}$ because the uniqueness theorem \cite{13,14}
that selects the Ashtekar -- Isham -- Lewandowski representation 
rests on the presence of the group Diff$^\omega(\sigma)$.
\item[ii)] {\it True Hamiltonian and Master Action}\\
Now what it is striking about (\ref{6.1}) is that it provides a {\bf true 
Hamiltonian}! Certainly we are only interested in the subset $\MC=0$
but nevertheless there is now only one constraint functional instead of 
infinitely many. It allows us to define the {\bf Master Action}
\be \label{6.2}
\MA_T=\frac{1}{\kappa}\int_0^T \;dt\{(\int_\sigma 
[\dot{A}_a^j E^a_j+A_t^j C_j])-\MC_E\}
\ee
where $C_j={\cal D}_a E^a_j$ is the Gauss constraint and $A_t^j$ a 
Lagrange multiplier. Actually we may also absorb the Gauss constraint
into the {\bf Master Constraint} giving rise to the {\bf Total Master
Constraint}
\be \label{6.1a}
\MC_T:=\frac{1}{2}\int_\sigma d^3x \frac{C(x)^2+q^{ab}(x) C_a(x) 
C_b(x)+\delta^{jk} C_j(x) C_k(x)}{\sqrt{det(q)}(x)}
\ee
which enables us to replace (\ref{6.2}) by the following expression without 
Lagrange multipliers 
\be \label{6.2a}
\MA_{T'}=\frac{1}{\kappa}\int_0^{T'} \;dt\{(\int_\sigma 
\dot{A}_a^j E^a_j)-\MC_T\}
\ee
Notice that the integrands of both (\ref{6.1}) and (\ref{6.1a})
are are densities of weight one 

The proof that 
the Ashtekar-Barbero phase space \cite{45a} with canonically conjugate 
coordinates $(A_a^j,E^a_j)$ modulo the the symmetries generated by 
the Gauss constraint on the constraint surface $C_j(x)=0,\;x\in\sigma$
is precisely the ADM phase space (see e.g. \cite{1} for a detailed proof) 
does not refer
to any dynamics, hence we may reduce (\ref{6.2}) and arrive at the
{\bf Master Action} in ADM coordinates 
\be \label{6.3}
\MA_T=\frac{1}{\kappa}\int_o^T \;dt\{(\int_\sigma 
\dot{q}_{ab} P^{ab})-\MC_E\}
\ee
in which there are no constraints anymore and $\MC$ is written in terms 
of the ADM Hamiltonian constraint $C(x)$ and the ADM diffeomorphism 
constraint $C_a(x)$. The equations of motion for $q_{ab}$ that follow from
(\ref{6.3}) are 
\be \label{6.4}
\dot{q}_{ab}=\frac{1}{\sqrt{det(q)}}[
\frac{2 P_{ab}-q_{ab} P^c_c}{\sqrt{\det(q)}}C+D_{(a} C_{b)}]
\ee
and can, in principle, be inverted for $P^{ab}$ off the constraint surface 
$\MC=0$ in order to invert the Legendre transformation. However, since the 
right hand side of (\ref{6.4}) contains second
spatial derivatives of $q_{ab},P^{ab}$ and moreover is a polynomial of 
third
order in $P^{ab}$, the functional 
$P^{ab}=F^{ab}[q,\partial q,\dot{q},\partial q,\partial^2 q]$ will be a 
non-local and non-linear expression, that is, a 
{\it non-local higher derivative theory of dynamical, spatial geometry 
on $\sigma$} ($\partial q$ denotes spatial derivatives) which is 
spatially diffeomorphism invariant! 

Since, however, the 
functional $F^{ab}$ can presumably not be extended to $\MC=0$, the 
Lagrangean formulation of the {\bf Master Action} (\ref{6.2}) or 
(\ref{6.3}) is presumably not very useful for path integral formulations,
hence we will stick with (\ref{6.2}). In fact, since the Gauss
constraint can be explicitly solved at the quantum level we can work
with the {\bf Extended Master Constraint} (\ref{6.1}) rather than the 
{\bf Total Master Constraint} (\ref{6.1a}). 
Notice also that the Lagrangean formulation of the {\bf Master Action}
is unlikely to have a manifestly spacetime covariant interpretation.
This is because the group of phase space symmetries generated by 
the Hamiltonian and Diffeomorphism constraint through their Hamiltonian 
flow are only very indirectly related to spacetime diffeomorphisms,
see \cite{1} for a detailed exposition of this relation.
\item[iii)] {\it Path Integral Formulation and Spin Foams}\\
The terminology {\it True Hamiltonian} is a little misleading because 
we are not really interested in the theory defined by the {\bf Master
Action} (\ref{6.2}). It appears only as an intermediate step in the 
path integral formulation of the theory. Here we will only sketch how this 
works, more details will follow in future publications \cite{54}.
For simplicity we describe the construction of the path integral for
the {\bf Extended Master Constraint} at the gauge invariant level, 
one could do it similarly with the {\it Simple Master Constraint}
used in previous sections at the gauge and spatially diffeomorphism 
invariant level as well, in which case one would use the space 
${\cal H}_{Diff}$ instead of ${\cal H}_{Kin}$ as a starting point.\\
\\
We assume to be given a self-adjoint {\bf Extended Master Constraint 
Operator}
$\MCO_E$ on the kinematical Hilbert space ${\cal H}_{Kin}$. Since that 
space
is not separable, we cannot follow the direct integral construction to
solve the constraint but if $\MCO_E$ is not graph changing then we can 
use the direct integral construction on separable subspaces of 
${\cal H}_{Kin}$,
see section \ref{s5}. In any case,
according to the general {\bf Master Constraint Programme} sketched in
section \ref{s2}, the constraint $\MCO=0$ can be solved, heuristically, by 
introducing the Rigging Map (see appendix)
\be \label{6.5}
\eta:\; \Phi_{Kin}\to (\Phi_{Kin}^\ast)_{Phys};\;\;f\mapsto 
\eta(f):=\lim_{T\to\infty} \int_{-T}^T\; \frac{dt}{2\pi}\;
<e^{it\MCO_E} f,.>_{Kin}
\ee
and the physical inner product 
\be \label{6.6}
<\eta(f),\eta(f')>_{Phys}:=\eta(f')[f]
=\lim_{T\to\infty} \int_{-T}^T\; \frac{dt}{2\pi}\;
<e^{it\MCO_E} f',f>_{Kin}
\ee
where $f,f'\in\Phi_{Kin}=\mbox{Cyl}^\infty$ are e.g. gauge invariant 
spin-network states. Notice that (\ref{6.6}) can formally be written as 
\be \label{6.7}
<\eta(f),\eta(f')>_{Phys}=<f',\delta(\MCO_E) f>_{Kin}
\ee
and defines a generalized projector which is of course the basic idea 
behind the RAQ Programme, see \cite{44} and references therein. 

Formula 
(\ref{6.7}) should be viewed in analogy to \cite{20} which tries to define 
a generalized projector of the form 
$\prod_{x\in\sigma} \delta(\hat{C}(x))$ at least formally where 
$\hat{C}(x)$ is the 
Hamiltonian constraint of \cite{2,3,4}. However, this is quite difficult
to turn into a technically clean procedure for several reasons: First
of all the $\hat{C}(x)$, while defined on ${\cal H}_{Kin}$ are not 
explicitly
known (they are known up to a diffeomorphism; they exist by the axiom of 
choice). Secondly they are not self-adjoint whence the exponential is 
defined at most on analytic vectors of ${\cal H}_{Kin}$. Thirdly, there 
is an infinite number of constraints and thus the generalized projector 
must involve a path integral over a suitable Lagrange multiplier $N$ and 
one is never sure which measure to choose for such an integral without
introducing anomalies. Fourthly and most 
seriously, the $\hat{C}(x)$ are not mutually commuting and since products 
of projections define a new projection if and only if the individual
projections commute, the formal object 
$\prod_{x\in\sigma} \delta(\hat{C}(x))$ is not even a (generalized)
projection. If one defines it somehow on diffeomorphism invariant states
(which might be possible because, while the individual $\hat{C}(x)$ 
are not diffeomorphism invariant, the product might be up to an 
(infinite) factor) then that problem could disappear because the 
commutator of two Hamiltonian constraints annihilates diffeomorphism
invariant states \cite{2,3,4}, however, this would be very hard to
prove rigorously. It is probably due to these difficulties and the 
non-manifest spacetime covariance of the amplititudes computed in 
\cite{20} for the Euclidean Hamiltonian constraint that the spin foam 
approach has chosen an alternative route that, however, has no clear 
connection with Hamiltonian formalism so far. 

Our proposal not only removes these four problems it also has the 
potential to combine the canonical and spin foam programme rigorously:\\
The ordinary amplitude $<f',e^{it\MCO_E} f>_{Kin}$ in (\ref{6.6}) 
should have a path integral
formulation by using a Feynman-Kac formula. More precisely, this amplitude 
should be defined as the analytic continuation $t\mapsto -it$ of the 
kernel underlying the {\bf Bounded Contraction Semigroup}
\be \label{6.8}
t\mapsto e^{-t\MCO_E},\;\;t\ge 0
\ee
relying on the {\bf positivity} of $\MCO_E$. Now for contraction 
semi-groups
there are powerful tools available, associated with the so-called 
{\bf Osterwalder -- Schrader} reconstruction theorem, that allow 
to connect the Hamiltonian
formulation with the a path integral formulation \cite{55,56} in terms 
of a probability measure $\mu$ on the space $\overline{\ab}$ of 
distributional connection {\bf histories}. 
Let us 
suppose that we can actually carry out such a programme then, relying
on the usual manipulations, 
the corresponding path integral should have the form (for $t\ge 0$)
\be \label{6.9} 
<f',e^{it\MCO_E} f>_{Kin}=
\int_{\overline{\ab}} d\mu(A) \overline{f'(A_t)} f(A_0)
=\int_{\overline{\ab}} d\nu(A) \overline{f'(A_t)} f(A_0)
\int_{\overline{\overline{{\cal E}}}} d\rho(E)
e^{i\MA_t(A,E)}
\ee
where $A_t$ denotes the point on the history of connection configurations 
at ``time" $t$ and $\nu,\rho$ is some measure on the space of 
distributional connection and 
electric field histories $\overline{\ab},\;
\overline{\overline{{\cal E}}}$ respectively,
both of which are to be determined (the 
first equality in (\ref{6.9}) is rigorous while the second is heuristic
but can be given meaning in an UV and IR regularization). 
The path 
integral of course is to be understood with the appropriate boundary
conditions.
Notice that in order to obtain the generalized projection we still have 
to integrate (\ref{6.9}) over $t\in \Rl$ which is the precisely the 
difference between a transition amplitude and a projection.

In order to get a feeling for what those measures $\nu,\rho$ could be,
we notice that the framework of coherent states is usually very powerful
in order to derive path integrals \cite{56a}. Hence we could use, as 
a trial, the coherent states developed in \cite{29,30,31,32,33} which
have the important overcompleteness property and thus can be used 
to provide suitable resolutions of unity when skeletonizing
$\exp(it\MCO_E)=\lim_{N\to\infty} [\exp(it\MCO_E/N)]^N$. But then, at 
least
on a fixed (spacetime) graph, we know that the ``measure'' 
$\nu\otimes\rho$ is 
related to the product of heat kernel measures \cite{56b} on non-compact 
spaces of generalized connections, one for each time step, based on the 
coherent state transform introduced by Hall \cite{56b}.

Alternatively, one could use resolutions 
of unity provided by holonomy and electric field eigenfunctions in which 
case the heat kernel measures, at each time step, are replaced by the 
Ashtekar-Lewandowski measure $\nu=\mu_{AL}$ \cite{12} times a discrete 
counting measure $\rho$
which sums over spin-network labels. Then, when performing the 
Fourier transform \cite{56d} with respect to the Ashtekar-Lewandowski 
measure from the connection representation to the 
electric field (or spin-network) representation one obtains amplitudes 
with pure counting measures, that is, {\bf a spin foam amplitude}.  
These issues are currently under investigation \cite{54}. 
\item[iv)] {\it Regge Calculus and Dynamical Triangulations}\\
This new form of a path integral approach to quantum gravity with a clear 
connection to the Hamiltonian framework and thus a clear physical 
interpretation
of {\it what exactly the path integral computes} could also be useful
for other path integral formulations of gravity such as Regge calculus 
\cite{57} and dynamical triangulations \cite{58} because, due to the 
positivity of $\MC_E$ the convergence of the path integral might be 
improved, the ``conformal divergence" could be absent. Of course, the 
function $\MC_E$ has flat directions but hopefully they have small enough 
measure in order to ensure convergence.
\item[v)] {\it Lattice Quantum Gravity and Supercomputers}\\
Finally it is worthwhile to point out that while we have been working with 
the continuum formulation throughout, the {\bf Master Constraint Programme}
easily specilizes to a {\bf Lattice Quantum Gravity} 
version, see e.g. \cite{34a}. 
Namely, we can just restrict the theory,
when graph-non-changingly defined,
to an arbitrary but fixed graph and study how the theory changes under 
coarsening
of the graph (background independent renormalization
\cite{61}). In background 
dependent theories such as QCD this is done 
in order to provide a gauge invariant UV and IR cut-off, in LQG, however, 
this should be viewed rather as a restriction of the UV finite theory to a 
subset of states and the renormalization is to be understood in the sense 
of Wilson, hence constructs an effective macroscopic theory from a given 
fundamental, microscopic one by integrating out degrees of freedom.
On the other hand, on a lattice both background dependent theories and 
LQG (when the constraints are treated in the usual way in the quantum 
theory) suffer from the same drawback, namely the destruction of continuum 
symmetries such as Poincar\'e invariance for background dependent theories
or spatial diffeomorphism invariance for LQG \cite{59}. However, this 
is no longer the case with the {\bf Master Constraint Programme}: By 
definition the symmetries follow from the {\bf Master Constraint Operator}
and are inherently discrete. They are defined by a single operator and not 
a multitude of them, hence there are simply no operators which must form 
an algebra. Notice that on the lattice one can certainly define a discrete
version of $\hat{C}_a(x)$ but the corresponding operator algebra does not 
close,
this is different with the {\bf Master Constraint Programme}, the algebra
consists of a single operator, hence the algebra trivially closes, it is 
Abelean.

The advantage of a lattice version is that it can be easily implemented on 
a {\bf supercomputer}, not more difficult than for QCD, although there 
will be new computer 
routines necessary in order to accomodate the non-polynomial structure 
of LQG. Notice that the {\bf Extended Master Constraint Programme} on the 
lattice can be compared to the method of \cite{25} in the sense 
that there are no (constraint algebraic) consistency problems even when
working at the discretized level. However, the two methods are quite 
different 
because the {\bf Master Constraint Programme} does not fix any Lagrange 
multipliers (in fact there are none). 
\end{itemize}
We finish this paper with the same warning as in the introduction: To the 
best of our knowledge, the 
{\bf Master Constraint Proposal} is an entirely new idea which has been 
barely tested in solvable model theories and hence one {\bf must test 
it in such model situations} in order to gain faith in it, to learn
about possible pitfalls etc. It is well possible that we overlooked some 
important fact which invalidates the whole idea or at least requires 
non-trivial modifications thereof. The author would like to learn about
such obstacles and tests in model theories (for instance it is 
conceivable that Loop Quantum Cosmology \cite{18} provides a fast and 
interesting test). On the other hand, we hope to have convinced the reader 
that the {\bf Phoenix Project}, that is, the {\bf Master Constraint 
Programme} applied to LQG, is an attractive proposal, designed to 
hopefully make 
progress with the quantum dynamics of LQG. Criticism, help and 
improvements by the reader are most welcome.\\ \\
\\
{\bf Acknowledgements}\\
\\
It is our pleasure to thank Bianca Dittrich for fruitful discussions about 
the contents of section \ref{s3}.

\begin{appendix}

\section{Direct Integral Decompositions and Rigging Maps}
\label{sa}

For the benefit of the reader more familiar with the theory of generalized
eigenvectors in order to solve constraints, we briefly sketch the 
connection with the direct integral approach below. We do this for a 
general theory defined on a Hilbert space ${\cal H}_{Kin}$.\\
\\
Let $\Phi_{Kin}\subset {\cal H}_{Kin}$ be a dense subspace equipped with a 
finer 
topology than the subspace topology induced from ${\cal H}_{Kin}$ and let 
$\Phi_{Kin}^\ast$ be the algebraic dual of $\Phi_{Kin}$ equipped with the 
topology of pointwise convergence (no continuity assumptions). An element 
$l\in \Phi_{Kin}^\ast$ is called a generalized eigenvector with eigenvalue 
$\lambda$ with respect 
to a closable operator $\MCO$ which together with its adjoint is densely 
defined on the (invariant) domain $\Phi_{Kin}$ provided that
\be \label{3.28}
\MCO' l=\lambda l\;\;\Leftrightarrow
l(\MCO^\dagger f)=\lambda l(f)\;\;\forall f\in\Phi_{Kin}
\ee
Here $\MCO'$ is called the dual representation on $\Phi_{Kin}^\ast$. The 
subspace
of generalized eigenvectors with given 
eigenvalue $\lambda$ is denoted by 
$\Phi_{Kin}^\ast(\lambda)\subset\Phi_{Kin}^\ast$ 
and $(\Phi_{Kin}^\ast)_{Phys}:=\Phi_{Kin}^\ast(0)$ is called the physical 
subspace. 

In this generality the concept of generalized eigenvectors does not require
${\cal H}_{Kin}$ to be separable which is an advantage. The disadvantage 
is that
$(\Phi_{Kin}^\ast)_{Phys}$ does not automatically come with an inner 
product. 
However, in fortunate cases there is a heuristic procedure known under 
the name ``Rigging Map" \cite{44}. We only consider the case at hand
and assume to be given a self-adjoint operator $\MCO$ on ${\cal H}_{Kin}$. 
The 
Rigging Map is defined as the antilinear operation
\be \label{3.29}
\eta:\;\Phi_{Kin}\to \Phi_{Phys}\subset 
(\Phi_{Kin}^\ast)_{Phys};\;f\mapsto
\int_\Rl\; dt <\hat{U}(t) f,.>_{Kin}
\ee
where 
\be \label{3.30}
<\eta(f'),\eta(f)>_{Phys}:=
[\eta(f')](f)=\int_\Rl\; \frac{dt}{2\pi}\; 
<\hat{U}(t) f',f>_{Kin}
\ee
It is clear that (\ref{3.29}) formally defines a physical generalized 
eigenvector by displaying the generalized eigenvector condition in 
the form 
\be \label{3.31}
l(\hat{U}(t)^\dagger f)=0\;\;\forall t\in\Rl,\;\; f\in \Phi_{Kin}
\ee
because the measure $dt$ is translation invariant and 
$t\mapsto \hat{U}(t)=e^{it\MCO}$
is a one-parameter unitary group. For the same reason, $<.,.>_{Phys}$
is a sesquilinear form. Now $\eta$ is said to be a Rigging map
provided that the sesquilinear form defined in (\ref{3.30}) is positive 
semidefinite (if not definite, divide by the null space and complete) and 
provided that $\hat{O}'\eta(.)=\eta(\hat{O})$. The latter condition is 
again easy to verify in our case for a strong Dirac observable (it 
ensures that $(\hat{O}')^\ast=(\hat{O}^\dagger)'$ where $(.)^\ast$ is the 
adjoint on ${\cal H}_{Phys}$, hence adjointness relations are induced 
from ${\cal H}_{Kin}$ to ${\cal H}_{phys}$).

In the case that ${\cal H}_{Kin}$ is separable we can show that the 
sesquilinear
form $<.,.>_{Phys}$ is actually already positive definite. Choose a direct
integral representation of ${\cal H}_{Kin}$ with respect to $\MCO$. Then
we know by the spectral theorem that the operator $\hat{U}(t)$ is 
represented on ${\cal H}^\oplus_{Kin}(\lambda)$ by multiplication by 
$e^{it\lambda}$, hence 
\ba \label{3.32}
<\eta(f),\eta(f')>_{Phys} &=& 
\int_\Rl\; \frac{dt}{2\pi} <\hat{U}(t) f',f>
\nonumber\\
&=& \int_\Rl\; \frac{dt}{2\pi}\int_{\Rl} d\nu(\lambda) 
<e^{i\lambda t}f'(\lambda),f(\lambda)>_{{\cal H}^\oplus_{Kin}(\lambda)}
\nonumber\\
&=& \int_{\Rl} d\nu(\lambda) \delta_{\Rl}(\lambda)
<f'(\lambda),f(\lambda)>_{{\cal H}^\oplus_{Kin}(\lambda)}
\nonumber\\
&=&
\nu(\delta)<f'(0),f(0)>_{{\cal H}^\oplus_{Kin}(0)}
\ea
The positive factor of proportionality is given by
\be \label{3.33}
\nu(\delta):=\lim_{\epsilon\to 0} \nu(\delta_\epsilon)
\ee
where $\delta_\epsilon$ is any family of smooth (thus measurable) functions 
converging to the $\delta-$distribution.

The framework of generalized eigenvectors can be connected even more 
precisely to the direct integral theory in the case at hand, at least
when ${\cal H}_{Kin}$ is separable, through the theory of 
{\it Rigged Hilbert Spaces} \cite{43}:\\
A Rigged Hilbert space is a Gel'fand triple 
$\Phi_{Kin}\hookrightarrow {\cal H}_{Kin}\hookrightarrow \Phi_{Kin}'$ 
consisting of a 
{\it nuclear space} $\Phi_{Kin}$, its {\it topological dual} $\Phi'_{Kin}$ 
(continuous
linear functionals) and a Hilbert space ${\cal H}_{Kin}$. The topologies 
of 
$\Phi_{Kin}$ and ${\cal H}_{Kin}$ are connected as follows:
\begin{Definition} \label{defa.1} ~~~~\\
i)\\
A countably Hilbert space $\Phi$ is 
a complete metric space whose topology is defined by a countable family of 
Hilbert spaces $\Phi_n,\;n=1,2,..$ whose scalar products $<.,.>_n$ are 
consistent in the following sense: First of all, $\Phi_n$ is the Cauchy
completion of $\Phi$ in the norm $||.||_n$. Then, for any $m,n$ it is 
required that if $(\phi_k)$ is both a $||.||_m$ convergent sequence and an 
$||.||_n$ Cauchy sequence in $\Phi$ then $(\phi_k)$ is also
$||.||_n$ convergent. We may w.l.g. assume that 
$||.||_n\le ||.||_{n+1}$ on $\Phi$. Then the metric on $\Phi$ is given by
\be \label{3.34}
d(\phi,\phi'):=\sum_{n=1}^\infty \;2^{-n} 
\frac{||\phi-\phi'||_n}{1+||\phi-\phi'||_n}
\ee
It is easy to verify that $\Phi=\cap_{n=1}^\infty \Phi_n$ and the inclusion
$\Phi_{n+1}\subset \Phi_n$ holds.\\
ii)\\
Let $\Phi'$ be the topological dual of $\Phi$ (continuous linear 
functionals) and $\Phi'_n$ the topological dual of $\Phi_n$. By the Riesz
lemma $\Phi'_n$ is isometric isomorphic with $\Phi_n$, that is, for any
$F\in \Phi'_n$ there is a unique element $\phi^{(n)}_F\in \Phi_n$ such 
that $F(\phi)=<\phi^{(n)}_F,\phi>_n$ for all $\phi\in \Phi_n$ and 
\be \label{3.35}
||F||_{-n}:=\sup_{0\not=\phi\in \Phi_n} \frac{|F(\phi)|}{||\phi||_n}
=||\phi^{(n)}_F||_n
\ee
Hence $\Phi'_n=:\Phi_{-n}$ can also be thought of as a Hilbert space.
Since $\Phi_{n+1}\subset\Phi_n$ any $F\in \Phi'_n$ is also a linear 
functional in on $\Phi_{n+1}$ and due to $||.||_n\le ||.||_{n+1}$ it is also
continuous. Hence we have the inclusion 
$\Phi_{-n}\subset \Phi_{-(n+1)}$ and it is easy to see that 
$\Phi'=\cup_{n=1}^\infty \Phi_{-n}$.\\
iii)\\
A Nuclear space $\Phi$ is a countably Hilbert space such that for each
$m$ there exists $n\ge m$ such that the natural injection 
\be \label{3.36}
T_{nm};\; \Phi_n\to \Phi_m;\;\psi\mapsto \psi
\ee
is a nuclear (that is, trace class) operator.\\
iv)\\
A Rigged Hilbert Space  
$\Phi\hookrightarrow {\cal H}\hookrightarrow \Phi'$ is given by a Nuclear
Space $\Phi$ and a Hilbert space $\cal H$ which is the Cauchy completion 
of $\Phi$ in yet another scalar product $<.,.>:=<.,.>_0$ such that 
if $\phi_k \to \phi$ in the topology of $\Phi$ then also $\phi_k\to \phi$
in the topology of $\cal H$. One can show that necessarily 
$\Phi_1\subset \Phi_0:={\cal H}=:\Phi'_0\subset \Phi'_1$ and hence we 
have an integer
labelled family of spaces with $\Phi_{n+1}\subset \Phi_n$.
\end{Definition}
The usefulness of the concept of a Rigged Hilbert space is that, given a
{\bf positive} self-adjoint operator $\MCO$ on a Hilbert space 
${\cal H}_{Kin}$,
a corresponding Rigged Hilbert space is often {\bf naturally provided as 
follows}:\\
Let $\cal D$ be a dense, invariant domain for $\MCO$, generically 
some space 
of smooth functions of compact support. Define positive sesquilinear forms
$<.,.>_n$ on $\cal D$ defined by
\be \label{3.37}
<\phi,\phi'>_n:=\sum_{k=0}^n <\phi,(\MCO)^k\phi'>
\ee
It is easy to see that $||.||_n\le ||.||_{n+1}$ and that the 
corresponding $\Phi_{Kin}=\cap_{n=1}^\infty \Phi_n$ (where $\Phi_n$ is the 
$||.||_n$ completion of $\cal D$) is a dense invariant domain for 
$\MCO$ and a countably Hilbert space. Whether it is also a Nuclear Space
depends on the operator $\MCO$, however, it is typically the case when 
$\MCO$ is a mixture of derivative and multiplication operators.

Thus, given a positive, s.a. operator $\MCO$ on a separable Hilbert space
${\cal H}_{Kin}$, a nuclear space $\Phi_{Kin}$ is often {\it naturally} 
provided.
\begin{Definition} \label{defa.2} ~~~\\
Let $\Phi'_{Kin}(\lambda)\subset\Phi'_{Kin}$ be the subspace of 
generalized 
eigenvectors with eigenvalue $\lambda$ in a Rigged Hilbert Space 
corresponding to a self-adjoint 
operator $\MCO$. For any $\phi\in\Phi_{Kin}$ and $\lambda\in\Rl$ we define 
an 
element $\tilde{\phi}_\lambda\in (\Phi'_{Kin}(\lambda))'$ by 
\be \label{3.38}
\tilde{\phi}_\lambda(F_\lambda):=F_\lambda(\phi)
\ee
for all $F_\lambda\in \Phi_\lambda'$. The map 
\be \label{3.39}
J:\;\Phi_{Kin}\to \cup_{\lambda\in \Rl} (\Phi'_{Kin}(\lambda))';\;
\phi\mapsto (\tilde{\phi}_\lambda)_{\lambda\in\Rl}
\ee
is called the generalized spectral resolution of $\phi\in\Phi_{Kin}$.

The operator $\MCO$ is said to have a complete set of generalized 
eigenvectors provided that the map $J$ in (\ref{3.39}) is an injection.
\end{Definition}
The motivation for this terminology is that $J$ is an injection if and
only if $\cup_{\lambda\in\Rl} \Phi'_\lambda$ separates the points of 
$\Phi_{Kin}$.
We notice that if $J(\phi)=(\tilde{\phi}_\lambda)_{\lambda\in\Rl}$ is the 
generalized spectral resolution of $\phi\in\Phi_{Kin}$ with respect to the 
operator 
$\MCO$ then $J(\MCO\phi)=(\lambda\tilde{\phi}_\lambda)_{\lambda\in\Rl}$
which suggests that there is a relation between the $\tilde{\phi}_\lambda$
and the direct integral representation 
$\phi=(\phi(\lambda))_{\lambda\in\Rl}$. This is indeed the case as the 
following theorem reveals.
\begin{Theorem} \label{th3.4} ~~~\\
A self-adjoint operator $\MCO$ on a separable Rigged Hilbert space
$\Phi_{Kin}\hookrightarrow{\cal H}_{Kin}\hookrightarrow\Phi'_{Kin}$ has a 
complete 
set of generalized eigenvectors corresponding to real eigenvalues.
More precisely:\\
\\
Let ${\cal H}_{Kin}=\int_\Rl^\oplus \; d\nu(\lambda)\; {\cal 
H}^\oplus_{Kin}(\lambda)$
be the direct integral representation of ${\cal H}_{Kin}$. 
There is an integer $n$ such that for $\nu-$a.a. $\lambda\in\Rl$ there is a 
trace class operator $T_\lambda:\;\Phi_n\to {\cal 
H}_{Kin}^\oplus(\lambda)$
which restricts to $\Phi_{Kin}$ and maps $\phi\in\Phi_{Kin}$ to its direct 
integral
representation $(\phi(\lambda))_{\lambda\in\Rl}$.
Then the map $J_\lambda:\;{\cal H}^\oplus_{Kin}(\lambda)\to 
\Phi'_{Kin}(\lambda)$
defined by $\xi\mapsto F^\xi_\lambda:=<T_\lambda^\dagger\xi,.>_n$ is a 
continuous, linear
injection and its image constitutes an already complete set of 
generalized
eigenvectors for $\MCO$. Restricting $\Phi'_{Kin}(\lambda)$ to the image 
of 
$J_\lambda$ and identifying $\xi,F^\xi_\lambda$, it 
follows form the Riesz lemma that the identity 
\be \label{3.40}
\tilde{\phi}_\lambda(F^\xi_\lambda)
=<\xi,\phi(\lambda)>_{{\cal H}^\oplus_{Kin}(\lambda)}
\ee
constitutes a one-to-one correspondence between $\tilde{\phi}_\lambda$
and $\phi(\lambda)$. Furthermore, combining $J_\lambda, T_\lambda$ we 
may identify (a dense set of) ${\cal H}^\oplus_{Kin}(\lambda)$ and (the 
subset 
defined by the image under $J_\lambda$ of) $\Phi'_{Kin}(\lambda)$ by
constructing $F^\phi_\lambda:=F^{T_\lambda\phi}_\lambda=
J_\lambda\circ T_\lambda\phi$. 
\end{Theorem}
The crucial part of this theorem is the existence of the nuclear operator
$T_\lambda$ for whose existence proof the machinery of Rigged Hilbert 
spaces is exploited. Theorem \ref{th3.4} gives a complete answer concerning
the question about  
the connection between generalized eigenvectors and the direct integral 
construction in the context of Rigged Hilbert Spaces and furthermore
guarantees that the physical Hilbert space 
${\cal H}_{Phys}:={\cal H}^\oplus_{Kin}(0)\cong\Phi'_{Kin}(0)$ is as large 
as mathematically possible, 
given $\MCO$. This does not mean, however, that it is as large as 
physically necessary.

\end{appendix}

\end{document}